%% file: main.tex
\documentclass[3p]{elsarticle}

\usepackage{tikz}
\usepackage{lmodern}
\usepackage{graphicx}
\usetikzlibrary{arrows.meta,
                chains,
                positioning,
                shapes.symbols,
                arrows,
                backgrounds}

\usepackage{amsmath}
\usepackage{amsfonts}
\usepackage{bibentry}
\usepackage{float}
\usepackage{subcaption}
\usepackage{microtype}
\usepackage{lineno}
\usepackage{ulem}
\usepackage[T1]{fontenc}
\usepackage{algorithm}
\usepackage{algpseudocode}
\usepackage{layouts}

\tikzset{
  mycross/.pic={
    \draw[pic actions] 
      (-3pt,0) -- (3pt,0)
      (0,-3pt) -- (0,3pt);
  },
}

\input{variables}

\begin{document}


\begin{frontmatter}

\title{Mean Mesh Adaptation for Efficient CFD Simulations with Operating Conditions Variability}
\author[1,2]{Hugo Dornier\corref{cor1}}
\ead{hugo.dornier@onera.fr}
\author[3]{Olivier P. Le Maître}
\author[2]{Pietro M. Congedo}
\author[1]{Itham Salah el Din}
\author[1]{Julien Marty}
\author[4]{Sébastien Bourasseau}
\address[1]{DAAA, ONERA, Institut Polytechnique de Paris, 92190 Meudon, France}
\address[2]{Inria, Centre de Math\'ematiques Appliqu\'ees, Ecole Polytechnique, IPP, Route de Saclay, 91128 Palaiseau Cedex, France}
\address[3]{CNRS, Centre de Math\'ematiques Appliqu\'ees, Ecole Polytechnique, IPP, Route de Saclay, 91128 Palaiseau Cedex, France }
\address[4]{{DAAA, ONERA, Institut Polytechnique de Paris, 92322 Ch\^atillon, France}}
\cortext[cor1]{Corresponding author}


\begin{abstract}

When numerically solving partial differential equations, for a given problem and operating condition, adaptive mesh refinement (AMR) has proven its efficiency to automatically build a discretization achieving a prescribed accuracy at low cost. However, with continuously varying operating conditions, such as those encountered in uncertainty quantification, adapting a mesh for each evaluated condition becomes complex and computationally expensive. To enable more effective error and cost control, this work introduces a novel approach to mesh adaptation. The method consists in building a unique adapted mesh that aims at minimizing the average error for a continuous set operating conditions. In the proposed implementation, this unique mesh is built iteratively, informed by an estimate of the local average error over a reduced set of sample conditions. The effectiveness and performance of the method are demonstrated on a one-dimensional Burgers equation and a two-dimensional Euler scramjet shocked flow configurations.

\end{abstract}


\end{frontmatter}

\section{Introduction}

In computational fluid dynamics (CFD), geometries under inflow and boundary conditions of interest can result in complex local phenomena and interactions, such as shock and flow separation patterns. An accurate evaluation of the quantities of interest (global performance, local metrics...) is conditioned by the capability of the calculation to capture these patterns with a control on the error. The latter is highly depending on the computational domain mesh discretization. As the regions of interest can appear in large zones as well as in very local ones, with strong gradients for example, complying to an accurate calculation of both without knowing \textit{a priori} where they should appear would require such a dense mesh that the effort of calculation would loose all its practicality. Furthermore, one may be interested in more than a single inflow condition, rendering even less achievable such an approach. 

A solution commonly proposed by the CFD community is to use adaptive mesh refinement (AMR,~\cite{alauzet&loseille,Habashi2000,Baker1997}) to adjust iteratively and automatically the local mesh density to the flow structures and provide meshes achieving a prescribed accuracy while limiting the computational cost. Several adaptation strategies and criteria have been developed over the years and applied to problems of variable complexity~\cite{alauzet_turbo,Vivarelli2018} .\\

The present work proposes to develop a new method to extend the concept of mesh adaptation to continuously variable operating conditions. This variability can be encountered in several categories of studies such as operating domain analysis where operating conditions are usually spread out and can lead to very different solutions. Uncertainty quantification also requires to handle conditional variability albeit usually with smaller variations and discrepancies. Nevertheless, both can be computationally demanding because of the numerous computations at different conditions they require and their accuracy is hard to control for all conditions. 

To tackle these complexities, several methods have been developed for mesh adaptation in the case of variable conditions. The standard approach consists in adapting the mesh for each condition to evaluate~\cite{Langenhove2018,Capriati2022} which enables an excellent error control. However, the cost of this approach can be high, depending on the refinement needs and method used. This approach also requires a fully automated and robust refinement tool, in particular for the selection of the refinement parameters that may be influenced by the particular condition. Therefore, defining a unique adapted mesh yielding a controlled error for any condition is desirable to avoid a systematic adaptation.
A straightforward approach consists in adapting the mesh to a unique representative condition, named nominal condition in the following, and to use it subsequently for all other conditions. This approach is limited and won't achieve low errors for conditions that yield solution structures different from the nominal one. 
An alternative method was proposed in~\cite{Palacios2012}, where an adjoint-based local sensitivity analysis is employed to enrich the refinement criterion at the nominal condition and build a unique goal-oriented adapted mesh for all conditions. In their work the authors demonstrated the effectiveness of the method in controlling the error in a limited neighborhood of the nominal condition. Nevertheless, because the refinement is only informed by the nominal condition and its sensitivity, the method fails when there are strong topological changes in the flow.
Another approach was proposed in~\cite{Barral} for anisotropic adaptation. In this approach, the metric of the next mesh is computed for several conditions and combined to get a unique metric value used to generate the next mesh. In practice, a metric intersection technique is employed. Compared to~\cite{Palacios2012} the method in~\cite{Barral} can prove to be more robust. However, the metric is built only on the fixed set of conditions tested, and the resulting mesh may not be suitable for other conditions.

As a result, none of these methods allows to control at the same time the cost and error level for studies with arbitrary continuously variables conditions. The present work proposes an alternative original methodology based on adaptive mesh refinement resulting in a single adapted mesh for multiple flow conditions to control cost and error, referred as the \textit{mean mesh adaptation method} or \MMA. We propose a strategy to build this unique adapted mesh to minimize the average error for a whole continuous set of conditions of interest with cost management. The iterative adaptation method developed uses an estimation of the local mean error from a sub-set of prescribed quadrature conditions to mark the cells for refinement. The local error estimator chosen for the study is a well referenced interpolation error estimator~\cite{Peraire1987,Habashi2000,Benard&al}, as it is reasonably assumed that the theoretical solution is not known.

In this paper, the \MMA\ methodology is described in detail along with the criteria used to evaluate its behavior and performance. Then, the effectiveness of the method is demonstrated on a one-dimensional parametric Burgers equation for which the exact theoretical solution is available; this problem allows to perform a complete statistical characterization of the resulting errors and their sensitivity to the adaptation method parameters. As a first step toward the use of this methodology for CFD problems, a two-dimensional supersonic scramjet inlet in an inviscid flow field~\cite{Kumar81} is then considered. This case is widely used in the mesh adaptation literature~\cite{CastroDiaz1997,Alauzet2003,Loseille&al} as it yields complex shock structure and interactions.

\section{Adaptive mesh for finite volume method} 
\input{fvm_amr}
\section{Mean mesh adaptation method}
\label{sub:mean_amr} 
\input{mean_amr}
\section{Burgers case}
\input{burgers}

\newpage
\section{Scramjet case}
\input{scramjet}

\newpage
\section{Conclusion}

A new approach, the mean mesh adaptation, has been developed in the context of continuous conditional variability and has been applied to error based adaptive mesh refinement. 
The method consists in building a unique mean mesh from the average estimated error of several sample conditions. 
It's evaluation on finite volume CFD test cases, in 1 and 2 dimensions, has shown the ability of the method to robustly control the average error at a reasonable cost. 
It has been observed that using a Monte Carlo renewed quadrature for the average error computation allowed to choose small and cost effective construction sets with minimal impact on the error. Moreover, this sampling strategy guarantees that the adapted mesh converges toward the desired mean mesh. 
Therefore, for reasonable error targets \MMA\ allows for significant cost reductions when compared to \UR\ and \SMA\ adaptations strategies and is not subject to error stagnation like \NMA. This makes it a good addition to the existing mesh adaptation method for variables conditions existing in the literature~\cite{Palacios2012,Langenhove2018,Barral}.

In addition, once a mean mesh is built for a given error target, it's use for condition evaluation is fast and straightforward without any meshing expertise requirements.
The \MMA\ method has also some implementation advantages, the average error level can be estimated on the fly without any additional expense and the method's structure is inherently parallel, making it well-suited for HPC frameworks. 

Nevertheless, \MMA\ still faces some challenges. Creating good initialization states for fast solver convergence is complex and impacts the method overall cost. Our investigations have shown that \MMA\ becomes less cost efficient when targeting very low average errors and when working with high conditional variability. 

Future work will be dedicated to the development of new methodologies that are more adapted to high conditional variability and high accuracy and that are not necessarily based on a single common mesh. To do so the construction of collections of adapted meshes from which to infer the best mesh for a given condition is considered.  Optimal transport based mesh adaptation is also investigated to allow for direct mesh adaptation at unknown conditions.

\section*{Acknowledgement}
This work was partially supported by the SONICE project which is a France Relance project co-funded by the French Civil Aviation Authority (DGAC). France Relance benefits from EU funding via the NextGenerationEU initiative.

\bibliographystyle{elsarticle-num-names}
\bibliography{biblio.bib}

\end{document}

%% file: variables.tex
\newcommand{\MMA}{MMA}
\newcommand{\SMA}{SMA}
\newcommand{\NMA}{NMA}
\newcommand{\UR}{UR}
\newcommand{\param}{p}
\newcommand{\mesh}{\mathcal{H}}
\newcommand{\cell}{c} 
\newcommand{\constrset}{\mathcal{P}_\text{c}}

\newcommand{\constrsize}{{N_\text{p}}}

\newcommand{\domain}{\Omega}
\newcommand{\sol}{u}
\newcommand{\err}{\varepsilon}
\newcommand{\errh}{\widehat{\varepsilon}}
\newcommand{\bigerr}{\mathcal{E}}
\newcommand{\bigerrint}{\bigerr^\mesh_{\text{int}}}
\newcommand{\bigerru}{\bigerr^\mesh_{\text{u}}}
\newcommand{\errt}{\bigerr_{\text{tol}}}

\newcommand{\Espp}[1]{\mathbb{E}_{\param}\left[{#1}\right]}

\newcommand{\Esppint}{\Espp{\bigerrint}}
\newcommand{\Esppu}{\Espp{\bigerru}}
\newcommand{\errinth}{\widehat{\bigerrint}}
\newcommand{\erruh}{\widehat{\bigerru}}

\newcommand{\evalsize}{N_\text{e}}

\newcommand{\meanid}{\text{\MMA}}
\newcommand{\detid}{\text{\SMA}}
\newcommand{\meancost}{T_\meanid}
\newcommand{\detcost}{T_\detid}
\newcommand{\reffrac}{\alpha}
\newcommand{\scaling}{\gamma}
\newcommand{\cfdscaling}{{\scaling_{\text{cfd}}}}
\newcommand{\adscaling}{{\scaling_{\text{ad}}}}
\newcommand{\meanits}{n_\meanid}
\newcommand{\ns}{N_0}
\newcommand{\initcost}{C_{\text{cfd}}}
\newcommand{\initcostad}{C_{\text{ad}}}
\newcommand{\cfdcost}{T_{\text{cfd}}}
\newcommand{\adcost}{T_{\text{ad}}}

\newcommand{\eff}{\eta}
\newcommand{\increase}{{G_\alpha}}

\definecolor{sblue}{HTML}{4682B4}
\definecolor{sred}{HTML}{FB3640}
\definecolor{sgreen}{HTML}{C1C544}

%% file: fvm_amr.tex
This section reviews the tools that will be used for the \MMA\ method proposed in the present work. 

\subsection{Finite Volume method}
In this work, we rely on Finite Volume (FV) methods for the discretization and resolution of the problem equations. We restrict ourselves to a bounded d-dimensional ($d\ge 1$) computational domain, $\domain$, delimited by its boundary $\partial\domain$. The solution $\sol(x)$ over this domain is approximated by discretizing $\Omega$ into a mesh $\mesh$ composed of non-overlapping cells $\cell_i\, ,i\in\{1,..,N\}$:
$$
	\Omega = \overline{\bigcup_{i=1}^N c_i}, \quad c_i \cap c_j = \emptyset,\,   \forall 1 \le i \ne j \le N.
$$ 
We choose a cell centered approach and denote $x_i$ the center of $\cell_i$. The discrete approximation of $\sol(x)$ on $\mesh$, denoted $\sol^{\mesh}(x)$, is defined from the values at the cells' center $\sol_i \doteq \sol^{\mesh}(x_i)$. The system of equations, usually non-linear, satisfied by the discrete solution $\sol^{\mesh}$ results from the discretization on the mesh of the model equations, \textit{e.g.} conservation laws. We assume that the discrete problem is well-posed, in the sense that there is a unique solution $\sol^h$ associated to a given mesh $\mesh$, and that $\sol^{\mesh}$ converges toward $\sol$ when the cells size $h$ goes to zero.
We remark that the \MMA\ method does not depend on a particular FV method and is readily applicable to other discretization approaches, \textit{e.g.} finite elements and finite differences.

\subsection{Adaptive Mesh Refinement}

We are interested in models involving complex physical phenomena that yield solutions $\sol$ with complex spatial structures. These solutions require a fine mesh to properly capture the structures of the field and achieve an approximation error $|\sol-\sol^{\mesh}|$ small enough. As the computational cost of $\sol^{\mesh}$ increases with the number of mesh cells $N$, that is when the cell size $h$ is reduced, a cost-accuracy trade-off question rises. Further, different areas of $\Omega$ may require different discretization efforts, and the mesh size must be adapted in space to provide the most suited aforementioned trade-off. Adaptive mesh refinement (AMR) strategies have been proposed to iteratively adapt the local mesh size to the solution and optimize the cost-accuracy trade-off. These methods are particularly appealing because they don't require a prior knowledge of the solution's structure and have demonstrated their capability to generate very effective adapted meshes, \textit{i.e.} with minimal number of computational cells to achieve a prescribed accuracy.

\subsubsection{AMR strategies}

An iterative mesh adaptation algorithms is usually based on the analysis, at each iteration, of the solution $\sol^{\mesh}$ computed on the current mesh. This analysis provides quantitative insights on the needed modifications to generate a new mesh with a better spatial discretization. The mesh adaptation strategies are generally classified in the literature into three categories \cite{Baker1997}:\\

\begin{itemize}
\item The r-adaptation~\cite{Gnoffo1982} consists in deforming the current mesh in order to optimize the position of the degrees of freedom according to the solution. In this case the number of elements in the mesh remains constant, so no additional computational cost due to any mesh refinement is to be foreseen. The main drawback of this approach is that it can yield large mesh deformations with accuracy degradation as a result and could also impact the numerical stability of the calculations.
\item The h and p-adaptations~\cite{Berger1985} change the number of degrees of freedom associated to the mesh. The h-adaptation consists in subdividing (isotropically or not) prescribed elements of the mesh according to a specified criterion. The criterion can also help coarsening the mesh by merging neighboring cells. The p-adaptation locally adapts the order of the discretization scheme according to the criterion. The p-method is mainly used with discontinuous Galerkin solvers~\cite{Kompenhans2016}.
\item The m-adaptation method generates a new mesh over the entire domain according to a metric derived from the current solution $\sol^{\mesh}$. The m-adaptation is the foundation of the anisotropic mesh adaptation methods~\cite{alauzet&loseille}.
\end{itemize}

All these adaptation methods rely on a criterion, or metric, to guide the adaptation process. 
We can distinguish several types of criteria, all derived from the solution on the current mesh.

The first category of criteria, known as \textit{feature-based}, is based on the solution structure. As its name shows, these criteria identify the structures of interest from the discrete solution analyses, for example: shocks, expansion fans \cite{Lovely&Haimes}, boundary and shear stress layers \cite{Haimes,deck,Zhou&al}, vortex, recirculation zones \cite{Liu2019}, or turbulent areas~\cite{Benard&al,Grenouilloux2022}. %

Another family of criteria uses a posteriori error estimations. The most common strategy is based on the interpolation error of $\sol^{\mesh}$ to estimate the local error in each computational element. These error based criteria have been in use since the 1980s and have proven their effectiveness~\cite{Lohner1985,Oden1987,Peraire1987,Habashi2000}. Other approaches for local error estimation have been proposed in the literature, such as those based on energy norms~\cite{Wu1990}.

Finally,  \textit{goal-oriented} criteria use an estimation of the error in user defined quantities of interests. In most cases, goal-oriented criteria use an adjoint method to estimate the local contribution of each computational element to the error in the quantity of interest (\textit{e.g.}, aerodynamic coefficients). The method was introduced into CFD by~\cite{Giles1999,Venditti2003}.

\subsubsection{Error based adaptation}
\label{sub:det_adapt} 

In this section, we focus on the adaptation of the mesh $\mesh$ for a specific condition $\param$. We consider an h-refinement strategy with a criterion based on the error estimation to manage the adaptation.
The objective is to minimize the global error induced by the discretization while controlling the number of elements $N$. The global error over $\domain$ relative to the problem's true solution $\sol$ is given by

\begin{equation}
	\bigerru = \int_{\domain} \|\sol(x) - \sol^{\mesh}(x)\| \ dx = \sum_{\cell_i\in\mesh} \int_{c_i} \|\sol(x) - \sol^{\mesh}(x)\|dx,
	\label{eq:det_crit}
\end{equation}
where $\|\cdot\|$ is a suitable norm. Finding the mesh $\mesh$ with the smallest number of computational elements and minimizing $\bigerru$ is a complex optimization problem usually untractable. Instead, we rely on an iterative adaptation process that refines some cells into smaller ones. At each iteration, the solution is computed on the current mesh and a subset of cells is selected for refinement using a flag $f_i \in\{0,1\}$ to decide if the cell $c_i$ should be refined ($f_i=1$) or not ($f_i=0$). To achieve an efficient reduction of the error $\bigerru$ a criterion is needed to decide of the flags value. The \MMA\ method proposed later is independent of the particular criterion but in the following we consider the case of a criterion based on an estimation of the local error, as introduced in the next section.

\subsubsection{Interpolation error estimator}\label{sec:int_error}

In order to set the flag value, a reliable and fast estimation of the local error is required. We consider standard interpolation error estimations~\cite{Peraire1987,Habashi2000} and stress that subsequent development can be extended to more advanced error estimation strategies. 

We start by the one-dimensional case. We assume $\sol$ to be sufficiently differentiable. We denote $\Pi_h \sol$ the interpolation of $\sol$ over the mesh $\mesh$ from the values at the cells' centers $x_i$. In this work, we use piecewise constant (over the cells) interpolation schemes. Using the fundamental theorem of calculus and the Hölder's inequality, the error between the continuous function $\sol$ and its piecewise constant interpolation satisfies, for $x\in c_i$, the inequality  
\begin{equation}\label{eq:int_error1}
	\left| \sol-\Pi_h \sol \right|(x) = \left | \int_{x_i}^{x} \sol'(\xi)d\xi\right| \le \max_{\xi\in[x_i,x]} |\sol'(\xi)| \left | \int_{x_i}^{x} d\xi \right | \le \frac{h_i}{2} \max_{\xi\in[x_i,x]} |\sol'(\xi)| .
\end{equation}
In~\eqref{eq:int_error1}, $h_i$ refers to the cell size.

In practice, $\sol$ is unknown and its derivative $\sol'$ must be approximated from the values at the computed solution $\sol^{\mesh}$. We define ${d\sol^{\mesh}}/{dx}$ as the (piecewise constant) derivative of a piecewise linear reconstruction of $\sol^{\mesh}$ over the cells centers. With this estimate of $\sol'$ we obtain the final estimation of the interpolation error $\err_i$ in the cell $c_i$: 
\begin{equation}
  \err_i \doteq {h_i}\left | \frac{d\sol^{\mesh}}{dx} \right |. \label{eq:int_estim}
\end{equation}
For higher order interpolation schemes, it is possible to derive similar interpolation error estimators, see~\cite{Habashi2000}. In general, the degree of the $n$-interpolation error estimate will be based on the $(n+1)$-th derivative of the solution.

In dimension $d>1$, the common practice is to consider one-dimensional interpolation estimators in different directions~\cite{Habashi2000} or to use the maximal value of the higher order derivatives of $u$ in each cell~\cite{Benard&al}. In the present work, we apply~\eqref{eq:int_estim} between a cell $c_i$ and its neighboring cells $c_j \in \mathcal N_i$ to compute $\err_{i,j}$ from~\eqref{eq:int_estim}. Then, we define the interpolation error estimation as
\begin{equation}\label{eq:err_int_gen}
	\err_i = \left( \sum_{c_j\in \mathcal N_i}  \left(\err_{ij}\right)^q\right)^{1/q}, \quad 
	\err_{i,j} = \frac{h_{ij}}{2} \left| \frac{d u^{\mesh}}{d\tau_{i,j}} \right|, 
\end{equation}
where $h_{i,j}$ and $\tau_{i,j}$ are respectively the distance and the direction between the two neighboring cell's centers.  
In practice, we use a L2-norm ($q=2$) in the result sections below.

\subsubsection{Mesh refinement}
\label{seq:mesh_ref}

The estimation of the interpolation error in~\eqref{eq:int_estim} and~\eqref{eq:err_int_gen} is used to estimate the error of $\sol^{\mesh}$ in~\eqref{eq:det_crit} through
\begin{equation}\label{eq:estim_error}
\bigerru = \sum_{c_i \in \mesh} \int_{c_i} \left \| u(x) - u^{\mesh}(x)  \right \| dx \approx
	\bigerrint = \sum_{c_i \in \mesh}  \int_{c_i} \err_i dx = \sum_{i=1}^N |c_i| \err_i,
\end{equation}
where $|c_i|$ is the cell's volume. In order to reduce the error, one wants to refine cells with highest contributions $|c_i| \err_i$ to the error. To this end these cells meant to be refined are identified and flagged. Specifically, we set $f_i = 1$ to the fraction $0< \reffrac <1$ of cells having the highest contributions, $f_i=0$ to the others, and proceed with the refinement.
The rationale for this setting of $f_i$ is that, for smooth solutions, the refinement will roughly reduce $\err_i$ by half while dividing the cell's size by half ($h_i$ in~\eqref{eq:int_estim} and $h_{i,j}$ in~\eqref{eq:err_int_gen}). Note that the increase in the number of cells compensates the reduction of the cell's volume.

For the sake of simplicity, we consider a standard approach for isotropic mesh refinement. It consists in the subdivision of the cells with $f_i=1$ into $2^d$ sub-cells. This approach to mesh adaptation results in a hierarchy of nested meshes that can be represented by a tree, providing a data structure which can be exploited to efficiently perform various operations, such as solution interpolation to generate initial guess of the solution on the new mesh. We rely in practice on the library \texttt{adaptCells} developed at ONERA~\cite{cassio2015} for the subdivision phase. The utility can also perform smoothing to limit the differences in neighboring cell's sizes. Note that this error based mesh refinement strategy is transposable to other mesh adaptation methods (such as anisotropic mesh adaptation) as are the subsequent development of section \ref{sub:mean_amr}.

%% file: mean_amr.tex
This section introduces the \textit{Mean Mesh Adaption} (MMA) method which aims at building a unique mesh for all considered input conditions, while controlling the resulting averaged error.

\subsection{Variable Conditions}

We consider the case of a model problem involving parameters which can vary. As an example, these parameters can be related to boundary conditions or model constants. In the present work the situations of variable model forms or variable domains problems are not considered. Further, we introduce a probabilistic framework to describe the parameters variability. Specifically, let $(\Theta,\mathcal{A},\mathbb{P})$ be a probability triplet, with $\Theta$ the set of events, $\mathcal A$ a sigma-algebra and $\mathbb P$ a measure of probability. Let $\param(\theta)$ be an input parameter (also called condition) associated to the uncertain variable $\theta\in\Theta$,  $P$ be the range of $\param(\theta)$ and $f_\param$ be the probability density function of $\param(\theta)$. As a consequence the model solution $\sol$ will also depend on the parameter $\param$ and will appear as $\sol(x,\param)$ in the following equations. 

A robust approach to controlling the error would aim at constructing the mesh $\mesh$ with minimal number of cells satisfying the constraint
\[
	\mathbb P\left( \bigerr^\mesh (\param) > \errt \right) < \epsilon,
\]
where $\bigerr^\mesh$ is either the true or approximate error for $p$, $\errt$ is an error tolerance level and $\epsilon>0$ is a prescribed maximum probability threshold. However, estimating this probability can be very costly, as it would require too many observations requiring the evaluation of many uncertain input conditions, especially for small values of $\epsilon$. Therefore, a more feasible objective is pursued: it consists in minimizing the number of cells to achieve a prescribed average error. This approach is less constraining than enforcing a maximal error with high probability, and can produce a mesh yielding for some conditions errors that are much larger than the averaged one. The management of the resulting error distribution will be examined in the results section to evaluate the relevance of this choice. 

\subsection{Mean Mesh Adaptation}

First of all the expectation of a random variable $g(\param)$ is defined as
$$
	\Espp{h} = \int_\Theta g(p(\theta)) d\mathbb P(\theta) = \int_{P} g(p) f_\param (\param) d\param.
$$
Extending~\eqref{eq:estim_error} to the variable parameter case, the average error becomes
\begin{equation}\label{eq:mean_error}
\Esppu = \sum_{c_i \in \mesh} \Espp{\int_{c_i} \left \| u(x,\cdot) - u^{\mesh}(x,\cdot)  \right \| dx} \approx  \Esppint = \sum_{i=1}^N |c_i|\, \Espp{\err_i (\cdot)},
\end{equation}
where $\err_i(\param)$ is given by the estimated error based on $\sol^{\mesh}(x,\param)$. Equation~\eqref{eq:mean_error} shows that the average error is given by the sum over all the cells of the averaged estimated error $\err_i(\param)$. We now detail the estimation of these averaged interpolation error and the resulting refinement strategy to minimize $\Esppint$ for the smallest number of cells.

In general, the average of $\err_i(\param)$ is not known in closed form. Therefore we propose to estimate $\Espp{\err_i}$ by means of quadratures of the form
\begin{equation}\label{eq:aver_estim_loc}
	\Espp{\err_i } \approx \errh_i \doteq \sum_{k=1}^\constrsize w_k \err_i (\param_k),
\end{equation}
where $(\param_k, w_k)_{k=1,\dots,\constrsize}$ are respectively the quadrature points and weights. In the following $\constrset$ represents the set of quadrature points. For the choice of the quadrature rule, Gauss quadratures~\cite{Abramowitz} have been selected and are convenient for standard densities $f_\param$ and limited $d_\param$, leveraging sparse tensorization of one-dimensional quadrature rules. Gauss-Legendre quadrature will be tested in the examples hereafter. Deterministic quadrature are limited by the curse-of-dimensionality, thus stochastic quadrature are also considered, namely Monte Carlo (MC) integration~\cite{Liu2004}. Another motivation for using MC integration is the possible lack of smoothness in $\err_i(\param)$ that can drastically deteriorate the convergence rate of the deterministic quadrature which would require an overwhelming number of points even in dimension 1.
Further, we remark that an accurate estimation of $\errh_i$ is not mandatory as the objective is to rank the cells from their potential error reduction when refined. In addition, the adaptation process being iterative, it is quite robust to imprecise ordering of the cells' average error. Indeed, missed cells still have the opportunity to be refined during subsequent iterations. In any case, we follow the same strategy as for the case of a unique condition, flagging cells with $|c_i| \errh_i$ value relatively high enough, according to the prescribed fraction $\reffrac$.


\subsection{Mean Mesh Adaptation algorithm}

The adaptation process has been developed following the \MMA\ method described above. The full \MMA\ algorithm is provided in Algorithm~\ref{alg:adapt_proc}.
\input{figures/adapt_algo}

Even though it has not been evaluated in the present work, the algorithm enables the selection of the quadrature points $\param_k$ at each adaptation iteration which allows to adapt dynamically the order of the Gauss quadrature. It also enables the change of the conditions of the MC quadrature points, even for a fixed $\constrsize$. Changing the MC points at each iteration has a very beneficial impact on the behavior of the \MMA\ method as it prevents the adaptation to get fixed on the errors associated with the conditions in $\constrset$, at the expense of the errors at other conditions. Randomizing the conditions in $\constrset$ at it each iteration also allows a reduction of $\constrsize$. This behavior has been investigated and presented in the results section.

When $\constrset$ is set, the solutions $\sol^{\mesh}(x,\param_k)$ are computed and collected to estimate the local average error $\errh_i$ using~\eqref{eq:aver_estim_loc}. 
In addition, the average error is estimated at each iteration using
\begin{equation}\label{eq:final_bigerr_estim}
	\Esppint \approx \errinth \doteq \sum_{c_i \in \mesh} \sum_{k=1}^\constrsize w_k |c_i|\, \err_i(p_k).  
\end{equation}
The adaptation process stops when the estimated average error falls under a certain user-defined tolerance value $\bigerr_{\textrm{tol}}$. Otherwise, the flag are set and we use \texttt{adaptCells} to refine the mesh accordingly. This step may include advanced features, such as mesh smoothing and body fitting node adjustment. 
In any cases, the framework is identical to the framework for the one condition adaptation, except for the evaluation of the $\constrsize$ solutions, possibly performed in parallel, and the averaging of their local error estimates. When the conditions in $\constrset$ are kept unchanged from an iteration to the next, one can store the corresponding solutions $\sol^{\mesh}(x,\param_k)$ to initialize the solve on the next adapted mesh. When, on the contrary, the conditions in $\constrset$ are resampled at every iteration a overhead may be induced by lacking the knowledge of the solutions on the previous mesh. To mitigate this overhead, we implemented an initialization strategy that initializes the solver with the closest (in terms of Euclidean distance in $P$) solution on the previous mesh.

\subsection{Performance evaluation}

As one of the main goals of this study is to control the mean error on continuous sets of conditions, the strategy developed to accurately compare errors is based on the computation of the \textit{a posteriori} true average  error. Extending on equation~\eqref{eq:mean_error} leads to the following formulation:

\begin{equation}\label{eq:true_err}
\Esppu = \sum_{c_i \in \mesh} \Espp{\int_{c_i} \left \| \sol(x,\cdot) - \sol^{\mesh}(x,\cdot)  \right \| dx} \approx \erruh \doteq \sum_{c_i \in \mesh} \sum_{k=1}^\constrsize|c_i|\,\left \| \sol(x,\cdot) - \sol^{\mesh}(x,\cdot)  \right \| \,  w_k.
\end{equation}

Here, $\erruh$ is the estimation with a quadrature of the true average error $\Esppu$.
The true average error is computed using accurate reference solutions $\sol$ that can be exact analytical solutions of the problem or discrete solutions on very fine grids (at least as deep as the evaluated grid). 
In practice, the true average error on a given mesh is estimated with $\erruh$ using a set of validation quadrature points of sufficient size for the measurement to be converged enough on every iteration's mesh compared to the \MMA\ construction quadrature. In this work, the quadrature used for the true average error evaluation is the Monte Carlo quadrature with uniform sampling and equal weights.


\subsection{Cost analysis}\label{sec:cost}

In this section, the goal is to provide simple computational cost estimates for the different adaptation methods studied, enabling a better understanding of the advantages and limitations of each. We will compare only the costs of the mean and deterministic methods, as nominal adaptation is a specific case of mean adaptation (with $\constrsize=1$).
We distinguish between two types of cost: evaluation cost and construction cost. The primary cost is the evaluation cost, which represents the actual computational expense required to compute the solution for a given condition with each method. This evaluation cost applies to each of the $\evalsize$ evaluations that must be performed. The secondary cost, construction cost, pertains to \MMA\ and represents the offline cost of building the mean mesh. For smaller values of $\evalsize$, construction cost is significant to the total cost of \MMA. However, as $\evalsize$ increases, this fixed cost becomes negligible. Therefore, as we are focused on the mean costs of both \MMA\ and standard specific mesh adaptation (\SMA) methods, we will neglect the construction cost and compare only evaluation costs

By examining only the evaluation costs, the ratio between $\meancost$ and $\detcost$ provides a benchmark for the potential cost advantage of the mean method. Specifically, as $\evalsize \to \infty$, if $\meancost < \detcost$, then the mean method can outperform the deterministic one, though only for a sufficiently large number of evaluations. Otherwise, the mean method will not surpass the deterministic method.

Finally, we introduce two additional notations. We denote the cost of solving the CFD problem for a given condition on a specific mesh as $\cfdcost$, and the cost of all intermediary adaptation operations (such as remeshing, criterion calculation, projection, etc.) as $\adcost$.

During any mesh adaptation the mesh size at iteration $k>0$, $N_k$, depends only on the initial mesh size and the refined fraction (fixed in our case). Thus, $N_k=\ns\prod_{i=1}^{k}\increase = N_0\increase^k$. Where the parameter $\increase$ is the increase in mesh size at a given iteration depending on the refined fraction $\reffrac$ with the following relation $\increase=(1+(2^d-1)\reffrac),d\ge1$. 
In this work we consider a  cost model of the form : $T=C.N^\gamma ,\,\gamma\ge1$. This formulation link a given iteration's CFD and adaptation costs ($T_{\text{cfd},k}$ and $T_{\text{ad},k}$) to the the iteration's mesh size $N_k$ and ultimately to the refined fraction. These costs are thus written:
\begin{equation}
	T_{\text{cfd},k} = \initcost \increase^{\cfdscaling.k}\quad T_{\text{ad},k} = \initcostad \increase^{\adscaling.k}.
\end{equation}
This form introduces the $\cfdscaling$  and  $\adscaling$ parameters to adjust the scaling performance. High values of $\scaling$ will model codes with bad scaling performances while value close to 1 will represent codes with perfect scalability. In addition the initial mesh size $N_0$ is implicitely taken into account in the cost model constants $\initcost$ and $\initcostad$.

Considering $\meanits(\errt)$ the number of adaptation iterations which are required to build a mean mesh suited for a prescribed error level $\errt$, the evaluation cost $\meancost(\errt)$ of the  mean mesh adaptation process can be written as follows:
\begin{equation}
	\meancost(\errt) = \initcost {\increase}^{\cfdscaling.\meanits(\errt)}.
\end{equation}
For \SMA\ the number of adaptation iterations depends on the condition considered and thus writes $n_{\detid}(\errt,p)$. The associated cost is :
\begin{equation}
	\detcost(\errt) = \initcost\Espp{\sum_{k=0}^{n_{\detid}(\errt,p)} {\increase}^{\cfdscaling.k}} + \initcostad\Espp{\sum_{k=0}^{n_{\detid}(\errt,p)} {\increase}^{\adscaling.k}}.
\end{equation}

To compare both methods performance we build an efficiency estimation from $\meancost(\errt)$ and $\detcost(\errt)$: 
\begin{equation}
    \eff(\errt) = \frac{\detcost(\errt)}{\meancost(\errt)}.
\end{equation}
Under the following hypothesis:
\begin{enumerate}
    \item The adaptation cost is small compared to the CFD cost : $\initcostad \ll \initcost$.
    \item The variations of $n_{\detid}(\errt,p)$ are small and : $\initcost\Espp{\sum_{k=0}^{n_{\detid}(\errt,p)} {\increase}^{\cfdscaling.k}} \approx \initcost\sum_{k=0}^{\overline{n_{\detid}(\errt)}} {\increase}^{\cfdscaling.k}$.
\end{enumerate}
The efficiency $\eff$ can be approximated in two parts as :
\begin{equation}
    \eff(\errt) \approx  \increase^{\cfdscaling(\overline{n_{\detid}(\errt)}-\meanits(\errt))} + \sum^{\overline{n_{\detid}(\errt)}-1}_{k=1} \increase^{\cfdscaling(k-\meanits(\errt))}.
\end{equation}

This formulation allows us to analyze the method's principal cost characteristics and their influence on $\eff$. Indeed, when $\meanits(\errt) \le \overline{n_{\detid}(\errt)}$ the first term is always greater than one, therefore $\eff(\errt)\ge1$ and \MMA\ is more efficient than \SMA. Conversely, when $\meanits(\errt) > \overline{n_{\detid}(\errt)}$ no direct conclusions can be made on $\eff$ as it's value will depend on the difference between the number of adaptation iterations of both methods and on $\cfdscaling$. Nevertheless, small $\overline{n_{\detid}(\errt)}$ compared to $\meanits(\errt)$ and big scaling $\cfdscaling$ will reduce efficiency. Finally, this estimation does not take into account adaptation costs $\adcost$ (hypothesis 1) which will have a positive impact on $\eff$ as it only increases $\detcost$.

%% file: figures/adapt_algo.tex
\begin{algorithm}
\caption{\MMA adaptation}\label{alg:adapt_proc}
\begin{algorithmic}
\Require $\mesh_0$, $\errt$
\State $j \gets 1$
\State converged $\gets$ False
\While{converged=False}
\State Set $\constrsize$ quadrature points and weights $(\param_k, w_k)_{k=1,\dots,\constrsize}$ 

\For{$k:=1$ to $\constrsize$}
    \State solve for $\sol^{\mesh_j}(\param_k)$ on $\mesh_j$
\EndFor
\State for each cell $c_i$ compute local average error $\errh_i$ \Comment{Use equation \eqref{eq:aver_estim_loc}}
\State compute average error, $\widehat{\bigerr^{\mesh_j}_{\text{int}}}$ \Comment{Use equation \eqref{eq:final_bigerr_estim}}

\If{$\widehat{\bigerr^{\mesh_j}_{\text{int}}} < \errt $}
\State converged $\gets$ True
\Else
    \State Set the flag $f_i$ \Comment{Use section \ref{seq:mesh_ref} strategy}
    \State Subdivide flagged cells, $\mesh_{j+1}$ \Comment{With \texttt{adaptCells}}
    \State $j \gets j+1$
\EndIf
\EndWhile
\end{algorithmic}
\end{algorithm}

%% file: burgers.tex
In the following sections the performance in terms of error and cost of the \MMA\ is evaluated and compared to uniform refinement as well as existing adaptation methods used in the context of continuously variable conditions. To this extend, the following adaptation strategies have been implemented to be compared to the \MMA\ method:
\begin{itemize}
    \item \textbf{Uniform refinement (\UR)} which subdivides all cells of the mesh at each adaptation iteration (which is equivalent to $\alpha=1$), the error convergence with this method should reflect the order of the numerical method employed, therefore it is used as a reference for error comparisons. 

    \item \textbf{Specific mesh adaptation (\SMA)} for which a specific adaptation for each condition to be evaluated is performed. With this method the error level is well controlled, thus it will be used as an error convergence best case scenario. 

    \item \textbf{Nominal mesh adaptation (\NMA)} which uses the mesh adapted for a single specific flow condition (referred as the nominal condition) and calls the resulting mesh to tackle any other condition calculations. While the cost of this method stays low, the quality of the results cannot by ensured for most conditions. Hence error levels higher than those of \SMA\ are expected.
\end{itemize}

\subsection{Case presentation}
\label{sub:cas} 
The generalized 1D Burger's unforced convection-diffusion equation has been selected as a relevant test case for it's analytical solution properties~\cite{Hopf1950}.
\begin{equation}
	\frac{d\sol}{dt} + \sol \frac{d\sol}{dx} - \nu \frac{d^2\sol}{dx^2} = 0,
\end{equation}

The computational domain and the following boundary conditions have been chosen:
\begin{equation}
	x\in[0;1], \quad \sol(x=0,t)=1, \quad \sol(x=1,t)=-1.
\end{equation}

\begin{figure}[h]
	\centering
    \scalebox{0.6667}{\includegraphics[]{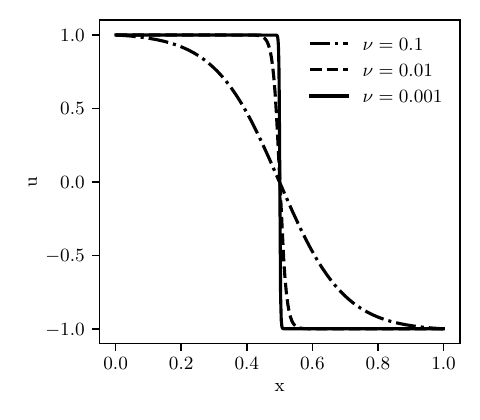}}
	\caption{Studied exact solution of the Burgers equation for different viscosity values. ($x_s=0.5$)}
	\label{fig:b_ref}
\end{figure}

With these conditions the monotone final solution highlights a strong gradient in a small zone around a position $x_s$, this size being controlled by the viscosity parameter, $\nu$ (see figure \ref{fig:b_ref}). The initial condition $\sol(x,t=0)$ is set using the following formulation:
\begin{equation}
\sol(x,t=0) = 
\begin{cases}
1 & \text{ if } x \le x_s \\ 
-1 & \text{ else }  
\end{cases}
\label{eq:c}
\end{equation}

In an infinite domain setting the strong gradient final position is precisely centered on $x_s$ at convergence. Here, because the computation is performed on a finite domain, a slight flux imbalance can be created at the domain boundaries convecting away from $x_s$ the strong gradient. As small values of $\nu$ will be considered in this work the time scale of this phenomenon is several orders of magnitude longer than the initial step diffusion, and is therefore negligible for the targeted convergence level.

The $\nu$ value chosen for the tests is 0.001 so as to generate a strong localized gradient in the computational domain. The problem is numerically solved using a cell centered fine volume method along with an implicit Euler temporal integration. The convective flux are discretized using a Roe scheme \cite{Roe1981} while the diffusive viscous flux are discretized with a centered scheme. Finally, the non-linear system is solved using the \textit{PETSc} library~\cite{petsc}.

To perform accurate error computations, the numerical solutions are compared to the analytical solution of the problem under the assumption that the strong gradient zone stays far from the domain's boundaries and that $\nu$ is small. This solution writes as follows:
\begin{equation}
	\sol(x,\param) =  -\text{tanh}\left (\frac{(x-x_\text{s})}{2\nu} \right ).
\end{equation}
In figures \ref{fig:cpgauss_b} we observe a consistent convergence of the solutions on uniformly refined grids toward the reference analytical solution. 
Finally, to compute the mean mesh criterion and validate the results, the positions of the strong gradient zone, $x_\text{s}$ are chosen uniformly spread between 0.4 and 0.6 following the law $x_\text{s} = 0.4+0.2\param,\,p\sim\mathcal{U}(0,1)$. The starting mesh for every adaptation is coarse with 25 uniformly distributed cells over the $[0,1]$ range.

\subsection{Results}

\subsubsection{Impact of the quadrature method}

\begin{figure}[htb!]

 \begin{subfigure}[t]{0.33\linewidth}
    \includegraphics[]{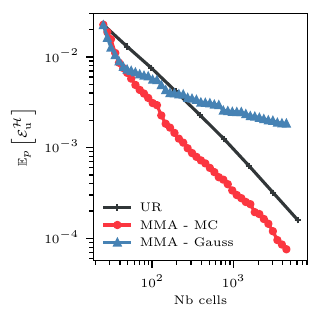}
    \caption{Average errors as functions of the number of cells in the mesh.}\label{fig:cpgauss_b}
 \end{subfigure}
 \begin{subfigure}[t]{0.33\linewidth}
  \includegraphics[]{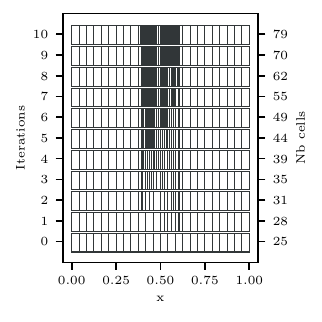}
  \caption{Grids with MC quadrature.}\label{fig:grids_burgers_mc}
\end{subfigure}
\begin{subfigure}[t]{0.33\linewidth}
  \includegraphics[]{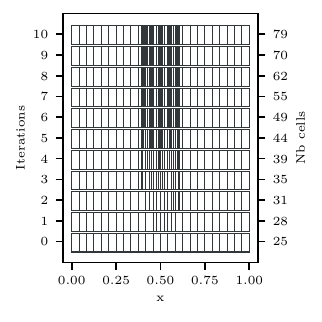}
  \caption{Grids with Gauss quadrature}\label{fig:grids_burgers_gauss}
\end{subfigure}
\caption{Comparison of average error and  mesh evolution with different \MMA\ quadratures (with $\reffrac=0.14$ and $\constrsize=5$).}
\end{figure}

As explained in the previous sections, the \MMA\ method requires to choose a quadrature to approximate the averaged error over the cells. 
In a first experiment, we compare the convergence of the average error with the number of cells in the mesh for quadrature sets of size $\constrsize=5$ and a refinement fraction $\reffrac=0.14$. The cases of Gauss and MC quadratures are contrasted, together with the evolution of the error for the \UR\ method. For the MC quadrature the random conditions $p_k$ are sampled using an uniform law with uniform weights $w_k=\frac{1}{\constrsize}$ 
The results are shown on figure~\ref{fig:cpgauss_b}. 

We first observe that the uniform refinement yields a first order convergence, as expected.
Further, we observe that initially the two \MMA\ variants perform better than the uniform refinement, with a slight advantage to the Gauss quadrature during the first few iterations. Subsequently, the MC quadrature continues to out-perform the uniform refinement, while the decrease of error is seen to slow down for the Gauss quadrature. Eventually, the averaged error for the Gauss quadrature becomes larger than for the uniform refinement, when the MC quadrature seems to converge at order one.

To understand this behavior figures~\ref{fig:grids_burgers_mc} and~\ref{fig:grids_burgers_gauss} illustrate the sequences of meshes generated by the two quadratures for the 10 first iterations. 
For both methods the refinement is located in the central area ($x\in [0.4,0.6]$) of the domain corresponding to the domain of variability of $x_\text{s}$ and where the steepest gradient of $\sol$ are expected. The MC quadrature which resamples the $\constrsize=5$ conditions at each iteration leads to a rather uniform refinement of this central area. In contrast, the Gauss quadrature keeps concentrating the refinement effort around the particular locations corresponding to the location of the maximum gradients ($x_\text{s}$) for the fixed $\constrsize=5$ Gauss points and disregards the refinement in between. This behavior is related to the limitation of the Gauss quadrature to accurately approximate discontinuous integrals. It is amplified when using lower refinement fraction values and by the fact that there is not constraints on neighboring cell size ratios. The MC quadrature would suffer from the same behavior if the quadrature conditions were not resampled at each iteration of the \MMA\ method.
Further, increasing $\constrsize$ delays the stagnation of the error for the Gauss quadrature (not shown), but does not improve much the error of the MC quadrature (see figure \ref{fig:errfull}). In fact, it can be seen that asymptotically, the MC quadrature needs about $1/5$ the number of cells of the uniform refinement to achieve a comparable error level, a ratio corresponding roughly to the fraction of the domain area requiring adaptation. 

From the finding of these experiments, we consider only the MC quadrature in the rest of the section.

\subsubsection{Performance of the adaptation approaches}

\begin{figure}[htb!]
    \begin{subfigure}[t]{0.49\linewidth}
	\centering
    \includegraphics[]{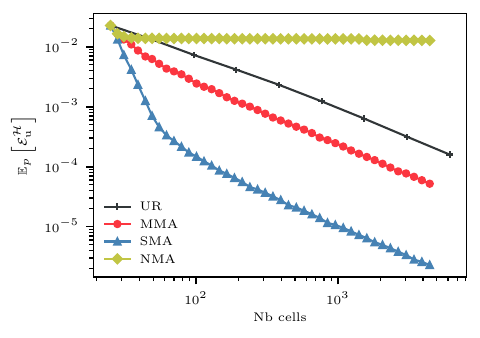}
	\caption{Evolution of the average error with number of cells.}\label{fig:errfull}
     \end{subfigure}
     \begin{subfigure}[t]{0.49\linewidth}
	\centering
    \includegraphics[]{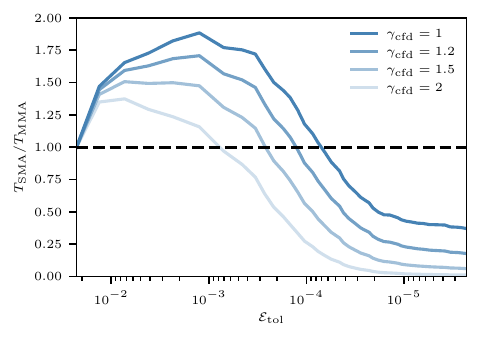}
    \caption{Evolution of the modeled evaluation cost efficiency of \MMA\ compared to \SMA\ as a function of error target}\label{fig:nmcomp}
	\end{subfigure}
    \caption{Comparison of uniform refinement, \MMA, \SMA\ and \NMA\ methods ($\reffrac=0.14$ and for \MMA\ $\constrsize=20$ and quadrature is MC).}
\end{figure}

Now that the quadrature for \MMA\ has been down-selected, its performance against other adaptation methods is investigated. Figure \ref{fig:errfull} shows the evolution of the average error during the adaptation process for the \MMA, \UR, \SMA\ and \NMA\ methods. The refined fraction $\reffrac$ is kept to $0.14$ and the number of quadrature conditions $\constrsize$ is set to 20 in order to ensure a good convergence and alleviate the small oscillations observed with $\constrsize=5$.

The error convergence trend remain the same when increasing $\constrsize$. Consequently, $\constrsize$ could be kept small to reduce construction cost for cost intensive cases. Nevertheless, with a greater value of $\constrsize$ the error convergence is smoother meaning that the error reduction is less influenced by individual quadrature conditions. Indeed, because the average error estimator is averaged over more quadrature conditions, the error variations induced by individuals have a smaller impact on the mean estimation.  These results foreshadow a good robustness of the method that will be investigated in detail in a next section.

With the \SMA\ method, the error is very significantly reduced at the beginning of the process, due to the small size of the zone of interest for each adaptation (one at a time). Then, a gradual return to order 1 decrease is observed as the proportion of cells in the zone of interest increases. To achieve an error level of $10^{-3}$, the method requires 50 cells in each of the meshes, which is five times less than for a mean mesh. However, this number needs to be considered in relation to the cost of such an approach, as each of the subsequent evaluations of conditions requires a complete mesh adaptation. Therefore, this method is not feasible for evaluating a large number of conditions (see next section).
Although the first two adaptation iterations of the \NMA\ method allow for an error reduction equivalent to the other methods, due to yet coarse meshes, the progressive specification of the mesh on a single operating condition limits the achievable progress and the average error reaches a plateau. Sharing a mesh among multiple conditions is therefore a significant source of error. Thus, the \NMA\ method, although inexpensive in a multi-condition context, has limited usefulness.
The issues identified with both of these approaches further highlight the importance of designing a low-cost and accurate adaptation method in a multi-condition context.

In addition to the previous error analysis, to better understand how the \MMA\ and \SMA\  methods compare, their computational cost is analyzed. 
Using the methodologies developed in the previous section the ratio between the cost per evaluation to reach a specific error level (efficiency) has been computed for the \MMA\ and  \SMA\ methods. For the Burgers case this ratio is represented against the error level on figure \ref{fig:nmcomp} for different solver scaling ($\cfdscaling$).
Here $\cfdscaling$ is chosen to vary between 1 and 2, 1 being the best case scenario as the convergence time per element is not increasing, while 2 represents the worst scaling scenario. The observations made with our 1D Burgers solver here would be close to a scaling of 1.
For errors levels above $5.10^{-5}$ the observed efficiency, for $\cfdscaling=1$, is greater than one. Therefore, if the objective is to make a mesh suited for a higher average error level, the cost of adapting the mesh for each condition will be higher than the cost of evaluating the solution for this condition on the equivalent mean mesh.
The position of the efficiency intersection point depends on the chosen adaptation parameters because $\reffrac$ directly influences the slope of the error convergence for the \SMA\ adaptation. The scaling $\cfdscaling$ also influences the intersection point position, as it gets bigger the majority of the cost gets concentrated on the biggest meshes, thus the cost of computing on the mean mesh gets higher than the cost of computing each of the smaller successive \SMA\ meshes. 

It is clear that the Burger case chosen here is very challenging for the \MMA\ process because of localized solutions and large studied range of conditions. Nevertheless, the method is capable of good cost reductions as long as the target error level is not to low and the solver scales well with the mesh size. In any cases, the method still benefits from its simplicity, as it allows to control the average error of the solutions without having to parametrize and run full adaptation processes once it is built. Finally, only the cost of the solution computation has been considered here without any overhead costs which could penalize \SMA\ compared to \MMA\ making the later a better choice.

\subsubsection{Robustness of \MMA} 

The convergence of $\Esppu$ is a key information in analyzing the quality of meshes resulting from \MMA\ process. Indeed, if the number of quadrature conditions evaluated to build an adapted mesh is high, then its performance should converge towards the ideal mean mesh (uniformly refined over the interval $[0.4;0.6]$). Thus, the role of $\constrsize$ is decisive, as the higher it is, the more representative the obtained mean mesh is. Yet, due to the nature of the mesh adaptation process with MC quadrature, the number of iterations also has a significant impact on the effective number of quadrature conditions. As $\constrset$ is changed at each adaptation iteration, the effective number of quadrature conditions is multiplied by the number of iterations. However, to achieve a target error level or a given number of cells, it is the refined fraction $\reffrac$ that determines the number of iterations. The effective number of quadrature conditions is in fact proportional to $\frac{\constrsize}{\ln(1+\reffrac)}$(Indeed, ${\constrsize}_{\text{eff}}=\constrsize.\frac{\ln(N_{\meanits})-\ln(\ns)}{\ln(1+\reffrac)} \propto \frac{\constrsize}{\ln(1+\reffrac)}$).
In order to quantify the impact of the effective number of quadrature conditions on \MMA\ and better understand its influence on the process robustness, a set of numerical experiments is conducted.  The adaptation processes are stopped when the refined mesh reaches a given size, $N=2500$ and the error on this mesh is recorded. In order to obtain comparable results, the mesh resulting from each adaptation process must be as close as possible to 2500 cells. This is ensured by determining the parameter $\reffrac$ based on a desired mean number of adaptation iterations $\meanits$. The resulting experimental design is therefore: $\constrsize\in[2;92]\,,\meanits\in[12;102]$ (the range of variation of $\reffrac$ is close to $[0.05;0.5]$). Additionally, the experiments are repeated 40 times in order to analyze the variability from one adaptation to another induced by the randomness of $\constrset$, for the same parametrization.

\begin{figure}[htb!]
	\centering
    \includegraphics[]{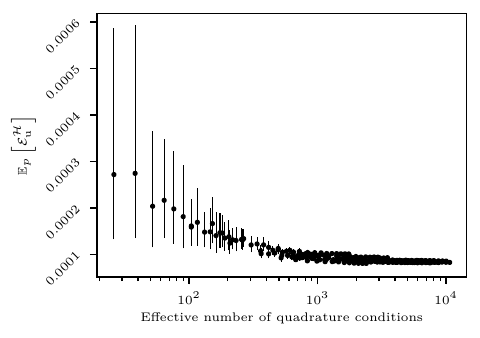}
	\caption{Average and 95\% confidence interval of the error on 40 realizations of the $\constrsize\in[2;92]\,,\meanits\in[12;102]$ design of experiments, using the \MMA\ method with MC quadrature}\label{fig:nbti}
\end{figure}
Figure \ref{fig:nbti} summarizes the average measured error over the 40 realizations of the experiments as a function of the effective number of quadrature conditions, as well as the associated 95\% confidence intervals.
On figure \ref{fig:nbti}, as the number of effective quadrature conditions increases, the average error converges towards a fixed value. At the same time, its confidence interval is also greatly reduced. With a small number of effective quadrature conditions, for example 24 (the lowest tested), a mesh of 2500 cells created by mean adaptation achieves, with 95\% confidence, an average error ranging from $1.1 \times 10^{-4}$ to $6 \times 10^{-4}$, an interval 120 times wider than for a high number of effective quadrature conditions (around 10000). Moreover, on average, the level of error reached when the number of effective quadrature conditions is low is 3 times higher than for a high number of effective quadrature conditions. Therefore, in addition to influencing the variability of the meshes obtained after the adaptation process, the number of effective quadrature conditions also impacts the achievable error level.

This result confirms the hypothesis formulated earlier, the average error tends towards a minimum value as the number of quadrature conditions increases. The meshes created under these conditions therefore tend towards a density that minimizes the error expectation and consequently towards the ideal average mesh, with a high level of confidence.
To design good mean meshes, one needs to maximize the number of effective quadrature conditions by choosing a very large value for $\constrsize$ and/or a very small value for $\reffrac$. However, drawing a large number of quadrature conditions involves a large number of computations and therefore a high cost to build the mesh, making \MMA\ difficult to apply.

\subsubsection{Use of interpolation error as an average error estimator}

\begin{figure}[htb!]
    \begin{subfigure}[t]{0.66\linewidth}
    \includegraphics[]{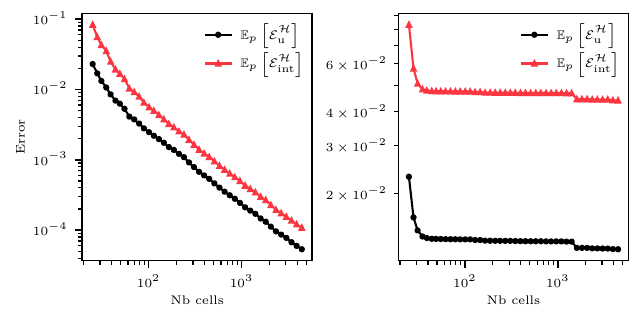}
	\caption{Evolution of the  true average error and it's estimation with the interpolation error for the \MMA\ (left) and \NMA\ (right) methods.}\label{fig:comp_err_itrp_mean_b}
    \end{subfigure}
    \begin{subfigure}[t]{0.32\linewidth}
    \includegraphics[]{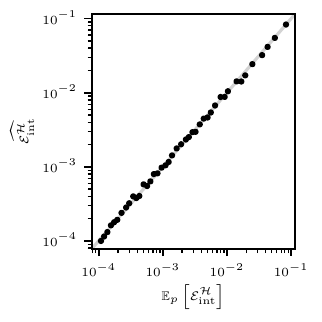}
    \caption{Comparison of the average interpolation error and its estimation on the quadrature conditions.}\label{fig:itrpcscv}
    \end{subfigure}
    \caption{Evaluation of the average error estimation (with  $\reffrac=0.14$, $\constrsize=20$ and MC quadrature).}
\end{figure}

As an analytical solution may not be available for the problem at hand, one would need to rely on the interpolation error estimator to evaluate the quality of the meshes. To ensure this approach wont misrepresent the error behavior,  error from the analytical solution $\Esppu$ has been compared to the interpolation error estimate $\Esppint$ for the \MMA\ and \NMA\ adaptation methods in figure \ref{fig:comp_err_itrp_mean_b}.

In the case of the \MMA\ method, the convergence for both errors is similar and the error estimated by interpolation behaves as an upper bound to the true value as could have been expected. This result shows that error in the Burgers solution considered is dominated by interpolation errors. Indeed, the shock position is well defined from the beginning of the adaptation (on the coarsest mesh) and won't drift significantly because of the boundary conditions.
Even in the unfavorable nominal case for which the refinement is not adapted to most conditions the interpolation errors dominate, thus the error estimated by interpolation bounds the true value.
For more complex applications, the interpolation error might not be as dominant, feature placement can vary with the resolution and can be interdependent and this approach could become misleading. At this stage no definitive conclusion can be drawn.

\subsubsection{Computation of the interpolation error on the quadrature points}

The previous results have shown that the interpolation error is a good, low cost, error estimate for the \MMA\  method. Nevertheless, the average error is still estimated on a large independent set that needs to be computed in addition to the quadrature conditions for validation studies. To alleviate this additional cost, the average error during the adaptation process needs to be directly computed on the successive quadrature sets conditions. In order to validate this approach, the average error using the interpolation error estimator has been computed on the validation set and on the successive quadrature sets. Both methods are compared on figure \ref{fig:itrpcscv}.

When performing the error measurement on the quadrature set no additional cost is required since the error is the same as the one used for refinement. Also, this allows to evaluate on the fly the solution's quality to inform a stop criterion for the adaptation. 
A strong agreement is observed, meaning that estimating the error on the quadrature set should not deteriorate the observations. This also emphasizes the fast convergence of the \MMA\ error estimator in the tested case for the purpose of building a flag function. Finally, because the refinement is performed after the error estimate computation during an adaptation step, there is no risk of contamination effects from refining the mesh for the same conditions as the ones used in the estimation.

%% file: scramjet.tex
In this section the \MMA\ method is analyzed on a higher complexity two dimensional fluid dynamics problem. A 2D scramjet inlet configuration. This device is meant to adapt supersonic and hypersonic upstream conditions and compress the flow field in order to create optimal conditions for a combustion chamber downstream. Thus, it operates at a supersonic regime generating complex shock interaction patterns.

\subsection{Case presentation}

\subsubsection{Case configuration}

\begin{figure}[htb!]
\begin{subfigure}[b]{0.5\linewidth}
\centering
\includegraphics[]{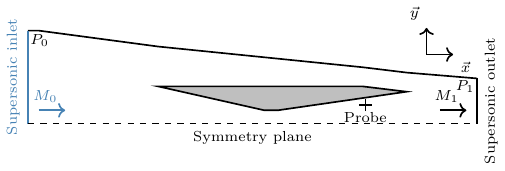}
\caption{Computational domain configuration.}\label{fig:scramjet}
\end{subfigure}
\begin{subfigure}[b]{0.5\linewidth}
\centering
\includegraphics[width=0.9\textwidth,trim={0 0px 0 0},clip]{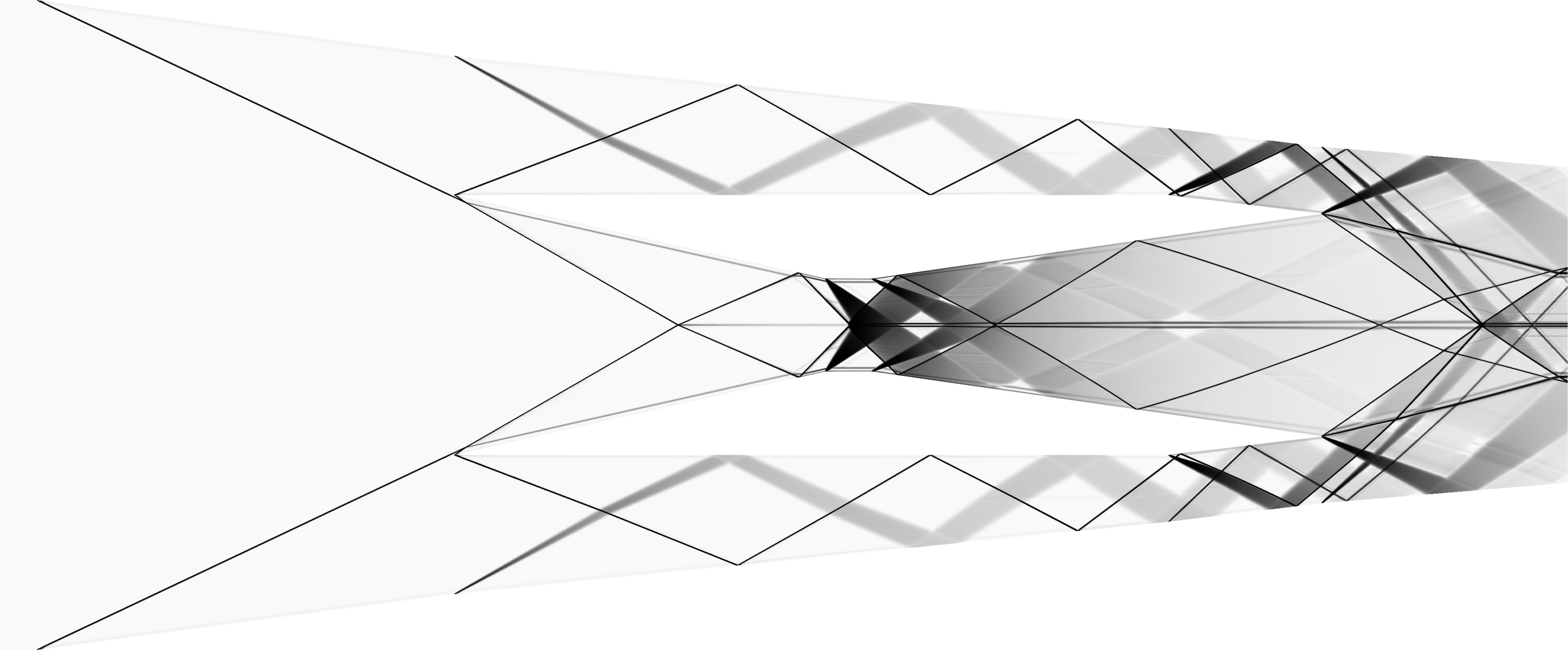}
\caption{Numerical Schlieren $M_0=3$.(symmetrized)}\label{fig:features}
\end{subfigure}
\caption{Scramjet problem.}
\end{figure}

The specific scramjet inlet configuration considered is a symmetric two-strut inlet originally introduced by \cite{Kumar81} in 1981 and widely used in the mesh adaptation literature \cite{CastroDiaz1997,Alauzet2003,Loseille&al,Langenhove2017}. Its geometry along with the computational domain configuration are presented on figure~\ref{fig:scramjet}.

The domain is divided into 4 parts: the upstream section, a central duct, two outer top and bottom (by symmetry) ducts and a downstream section where the duct flows merge. In addition a probing point is placed in the diverging section of the central duct to monitor the flow.

The continuous parameter that is chosen  to adjust the flow conditions is the inlet Mach number $M_0$. It affects the inflow condition and fixes the shock structures position and interactions in the domain.

\subsubsection{Operating conditions}

Shock boundary layer interactions and flow separations are common in scramjet flows and can have a huge impact on its behavior. Nevertheless, to focus the study on shocks  resolution, the scramjet's flow will be modeled using Euler equations closed for a perfect gas. For the whole study  the gas coefficient is fixed to $1.4$. At the nominal condition the flow enters the domain at $M_0=3$ and creates oblique shocks, expansion fans and complex interactions in both the central and outer ducts. as can be seen on figure \ref{fig:features}. Symmetric oblique shocks are induced by the upstream section reduction, and meets the struts leading edge. In the outer ducts their reflection bounces on the walls up to the outlet. The successive section modifications on these ducts create several expansion fans and oblique shocks that interact with each other in the downstream section.
In the central duct the upstream oblique shocks meets and reflect in the converging section. At throat new expansion fans and shocks are created resulting in interacting reflected structures in the diverging section and the downstream section.
As can be seen on figure \ref{fig:features} the union of the central and outer flows along with trailing edge structures make the downstream section flow highly complex and interdependent.

The range of Mach number considered in this study ranges from $M_0=2.9$ to $M_0=3.3$. The lowest bound has been chosen to avoid the formation of a strong shock at the central duct throat which results in a significant flow regime change.

The error measurements to evaluate the adaptation methods on this case are computed using a thin uniform reference mesh since there is no analytical solution available. This mesh has been uniformly subdivided six times (depth of 6) to reach a size around 6M cells. This size has been chosen mainly for computational cost reason. For the error measurements to be consistent, the cell size of the meshes that are being evaluated must remain smaller than the cells size of the reference, so that all meshes are embedded in the reference mesh. Therefore, in whats follows, the depth of all adaptations is limited to 6 levels (equivalent to the reference). 

\subsubsection{Flow solver}

The Euler equations for the case are solved by \textit{elsA} \cite{elsa2016} a finite volume and cell centered solver (ONERA-SAFRAN property). For the finite volume discretization of the equations the Roe scheme \cite{Roe1981} is used, it is well suited for configurations with shocks since it creates minimal numerical dissipation. While the cases are solved using an implicit backward Euler time integration and a LU-SSOR type linear solver.
The mesh adaptation strategy used in this paper works only on unstructured grids, therefore all the calculations are performed using \textit{elsA}'s unstructured capabilities. 

\subsection{Results} 

\subsubsection{Local error and mesh refinement}

\begin{figure}[htb!]
    \begin{subfigure}[t]{0.49\linewidth}
	\includegraphics[width=\textwidth]{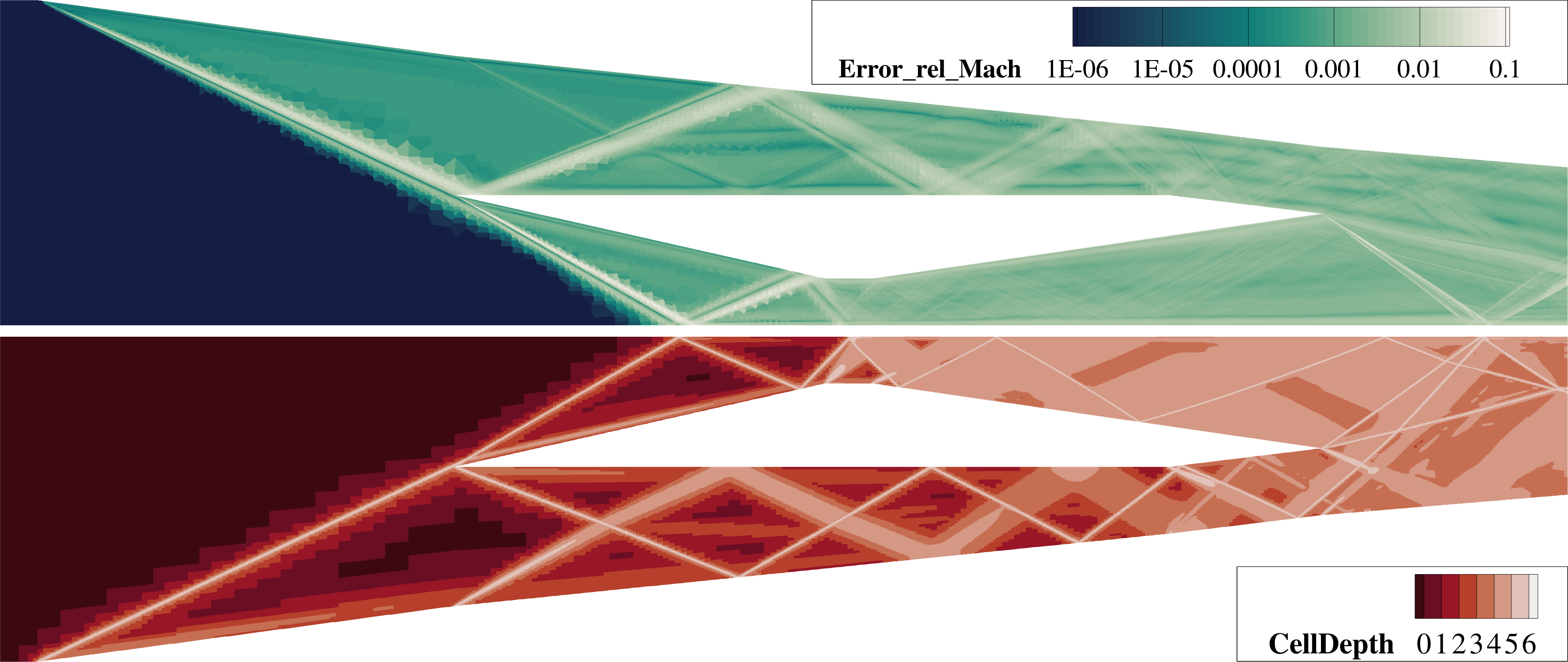}
	\caption{Nominal adapted mesh for $M_0=3$ (384K cells): local average error (top) and refinement depth (bottom).}\label{fig:local_err_det_s}
    \end{subfigure}
    \begin{subfigure}[t]{0.49\linewidth}
	\includegraphics[width=\textwidth]{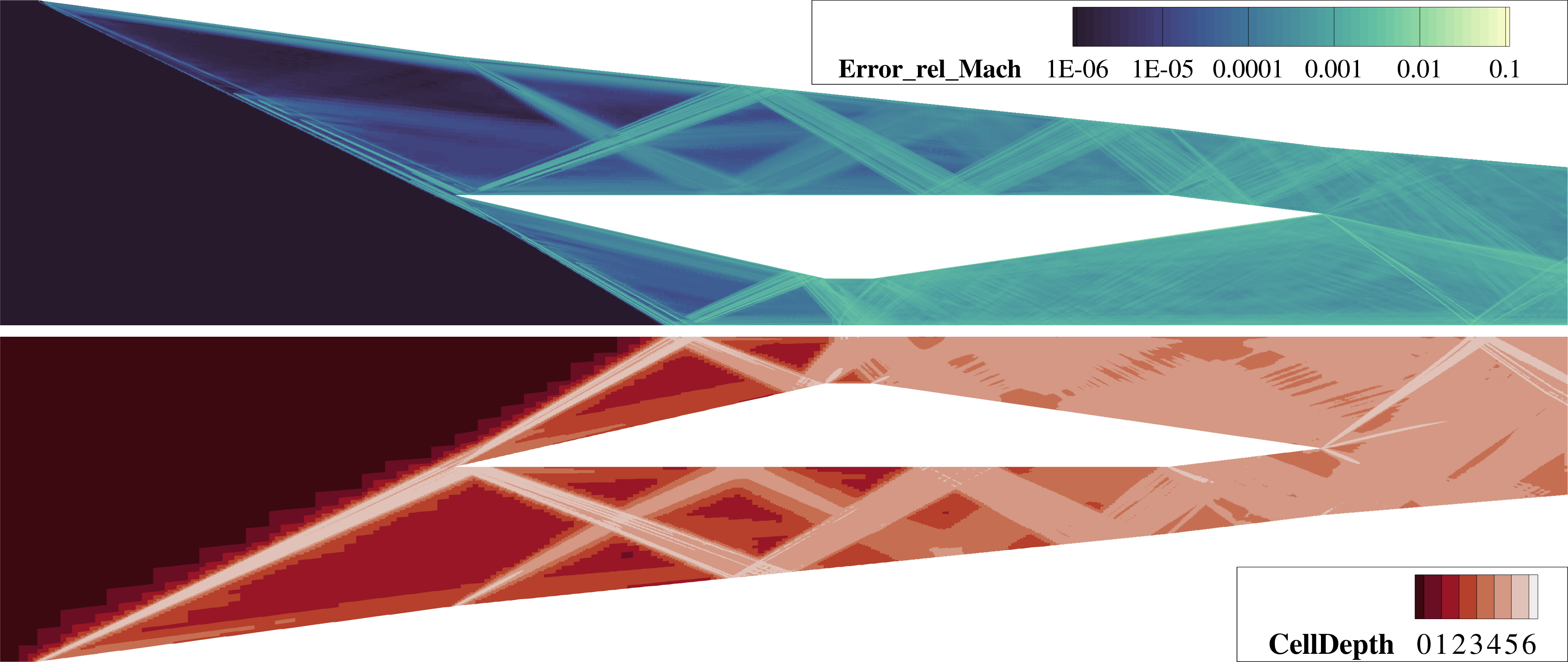}
	\caption{\MMA\ mesh (470K cells): local average error (top) and mesh refinement depth (bottom).}\label{fig:local_err_mean_s}
    \end{subfigure}
    \caption{Comparison of the meshes and local average errors for the \NMA\ and \MMA\ methods. The adaptations use 10 iterations from the initial mesh and $\reffrac=0.3$ (for \MMA\ $\constrsize=20$ and the quadrature is MC).}
\end{figure}

Before presenting in details the performances and characteristics of the newly developed \MMA\ method, and in order to better understand the accuracy advantages of \MMA\ compared to \NMA, the local error relative to the nominal inlet Mach number along with the refinement depth have been plotted on figures~\ref{fig:local_err_det_s} and \ref{fig:local_err_mean_s} for the final meshes of each method.

The figure \ref{fig:local_err_det_s} highlights the important discrepancies in the behavior of the \MMA\ and \NMA\ methods. With the \NMA\ method the error reaches up to 10\% of the nominal value in the upstream oblique shock region and converging section of the central duct. It is consistently above 1\% in most of the central and outer ducts as well as in the downstream section. Localized lower error values in this high error regions are due the deeply refined features of the nominal condition (especially on the upstream oblique shock). \NMA\ especially fails in the central duct diverging section and the downstream section where the flow topology undergoes drastic changes. In this area the refinement is almost as deep as with \MMA\ meaning that the error here comes mainly from the misplacement of the flow features induced by upstream shortcomings of the mesh. 
For these reasons, it is clear that the \NMA\ method is unable to handle important conditional variabilities and is not performing for flow topologies far from the nominal one. 
The error map resulting from \MMA\ also highlights some interesting features. First, the upstream section error level is close to zero as its uniformity leads to expect, resulting in low refinement depths. Then, low error levels are measured in the upstream oblique shock region since this shock is not very sensitive to the flow conditions, fine refinement has been possible in the regions it covers (see figure \ref{fig:local_err_mean_s}) resulting in good error control. In the outer ducts, a refined zone is observed around the reflections of the upstream oblique shock and around the reflections of an expansion fan. With the successive reflections, the refinement band from the oblique shock gets larger, covering an increasingly large area of variation of the features in this region.
Overall, where the sensitivity of the phenomena to the flow conditions increase, the measured error level gets higher. Especially in the downstream and the central duct diverging sections. In these regions the mesh depth is almost uniform and has not yet reached the same level as in the upstream shock region. The error is well managed by the \MMA\ method and the relative error does not get higher than 1\%. It is also relatively uniform as no strong error variation are observed on the challenging central duct and downstream regions. 

Additionally, there are complementary footprint on the mesh depth and error map left by the random sampling of the previous adaptation iterations. Especially after the first reflection of the upstream oblique shock. The over-representation of specific conditions at this location indicates a lack of convergence of $\errh_i$ at the last iteration. Since, the quadrature set is renewed at each adaptation iteration, this lack of regularity is meant to gradually disappear.

Finally, the sensitivity of the local phenomena with respect to the variable conditions seems to be an important aspect driving the \MMA\ performance. Indeed, where numerous phenomena sensitive to the varying conditions are known to occur, like in the central duct diverging section and the downstream section, to control the average error, most of the cells must be refined, resulting in almost uniform refinement. Whereas in the outer ducts and the upstream section the area with strong average error are localized and the refinement can be non-uniform. The main benefits brought by \MMA\ on this configuration are therefore upstream of the domain and the outer ducts where the refinement remains localized. In general, the \MMA\ method performance compared to \UR\ is conditioned by the overall sensitivity of the flow to the conditional variability imposed.

\subsubsection{Impact of the quadrature method}

\begin{figure}[htb!]
    \begin{subfigure}[t]{0.49\linewidth}
    \includegraphics[]{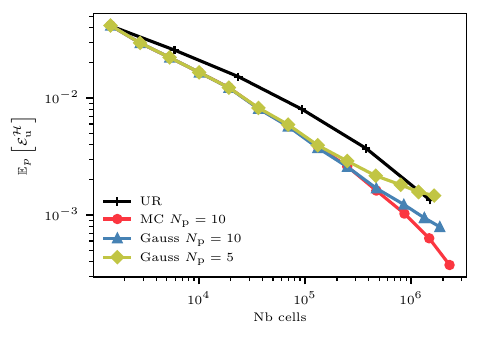}
    \caption{Average error for the different \MMA\ quadratures  (with $\reffrac=0.3$).}\label{fig:gauss_vs_random_s}
    \end{subfigure}
    \begin{subfigure}[t]{0.49\linewidth}

    \includegraphics[]{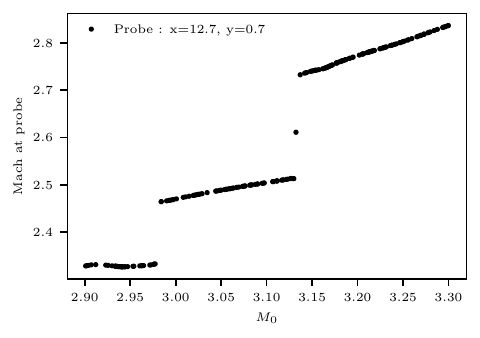}
	\caption{Local Mach at the probe point as a function of the inflow Mach number.}\label{fig:local_mach_probe}
    \end{subfigure}
    \caption{Influence of the \MMA\ quadrature and local Mach number regularity.}
\end{figure}

In 1D, the Burgers case analysis has shown that using a Gauss quadrature leads to error stagnation due to over-adaptation on the sample conditions, as there is no sampling update at each adaptation iteration. In order to verify this behavior on a higher dimension and a more complex case, the error convergence on the scramjet has been reported on figure \ref{fig:gauss_vs_random_s} for the MC quadrature (with $\constrsize=10$) and the Gauss quadrature (with $\constrsize=5$ and $\constrsize=10$), along with the \UR\ results.

Using Gauss and MC sampling yields comparable error levels until the $9^{th}$ step for $\constrsize=10$ and the $6^{th}$ step for $\constrsize=5$, after this point the error convergence is significantly slowed down for Gauss sampling. The nature of the local solution evolution when the condition change, plotted at the probe point (see fig \ref{fig:local_mach_probe}) explains this behavior. We can observe that the shocks in the domain are localized phenomena that induce steep Mach number variations when $M_0$ changes. Because of that the interpolation error is composed of spikes, impossible to accurately and reliably capture with a Gauss sampling. In this case the large value for $\reffrac$ chosen and the small position variation of some important features has allowed the Gauss sampling to keep up with the MC sampling longer than in the Burgers case.

On the other hand, not changing the quadrature set at each iteration with the Gauss sampling has resulted in significantly better initialization states and faster computational convergence. That makes it an interesting method to reach less demanding error levels faster.
As we are interested in the robustness of the method, MC sampling is kept for the rest of the scramjet study.

\begin{figure}[H]
	\centering
    \includegraphics[]{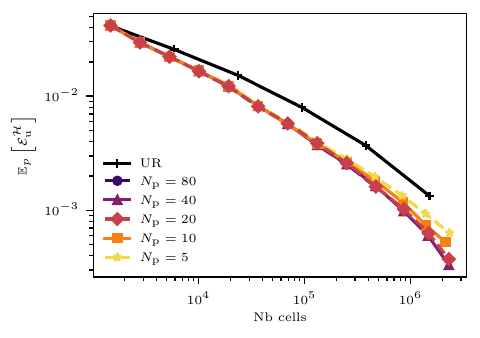}
	\caption{Average errors for different quadrature set size $\constrsize$ (with $\reffrac=0.3$ and MC quadrature ).}
	\label{fig:ns_s}
\end{figure}

The size of the initial mesh for the scramjet (1500 cells) is of the same order as the size of the last adapted meshes from the Burgers case. Also, because the case is 2D and the adaptation isotropic and hierarchical, each cell is subdivided into 4 cells (instead of 2) when flagged. Therefore, the cost of the refinement is higher making it even more critical to find the best compromise between cost and accuracy when choosing $\constrsize$. That's why the error convergence from five different values of $\constrsize$ have been compared on the figure \ref{fig:ns_s}. 

Overall, all the configurations tested give close results with limited discrepancies for the last adaptions with $\constrsize=5$ and $\constrsize=10$. It should be noticed that the sensitivity to $\constrsize$ is negligible between 40 and 80 as curves are superimposed. 
The high $\reffrac$ value chosen along with the good robustness of the random adaptation process  make the required number of effective quadrature conditions lower for this configuration.
In the following, the number of quadrature conditions has been fixed to $\constrsize=20$, in order to keep the cost under control while achieving high quality mean meshes.

\subsubsection{Performance of the adaptation approaches}

\begin{figure}[htb!]
    \begin{subfigure}[t]{0.49\linewidth}
	\centering
    \includegraphics[]{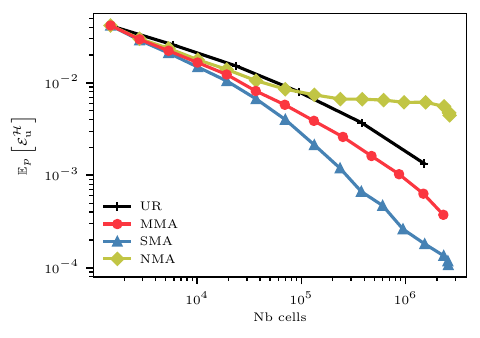}
	\caption{Evolution of the average error with number of cells.}\label{fig:comp_methods_s}
    \end{subfigure}
    \begin{subfigure}[t]{0.49\linewidth}
    \centering
    \includegraphics[]{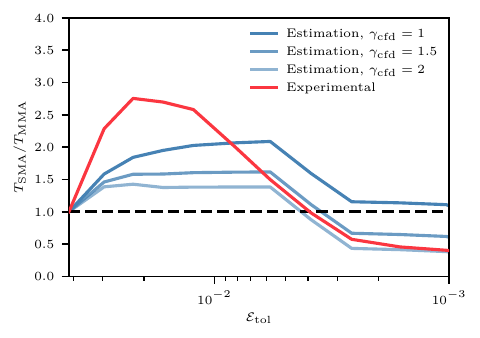}
	\caption{Evolution of the modeled and measured evaluation cost efficiency of \MMA\ compared to SMA\ as a function of error target}\label{fig:nmcomp_s}
    \end{subfigure}
    \caption{Comparison of \UR, \MMA, SMA\ and \NMA\ methods ($\reffrac=0.3$ and for \MMA\ $\constrsize=20$ and quadrature is MC).}
\end{figure}

The performance of the \MMA\ method is compared to the \UR, \SMA\ and \NMA\ methods similarly to the 1D Burgers study. The figure \ref{fig:comp_methods_s} shows the evolution of the average error, with $\constrsize=20$ and $\reffrac=0.3$ for each of these methods.

The worst performing method remains \NMA. During the first iterations it leads to similar error reduction levels as the other methods but from the fifth iteration the error reaches a plateau almost until the last iteration. This plateau is due to the specification of the mesh for the nominal condition ($M_0=3$).  The depth of the mesh is limited on the studied case. Therefore, as more and more high error cells get to the maximum depth, the amount of refined cells decrease. This amount ultimately goes to zero when 30\% (the refined fraction chosen) of the biggest error contributors get to the maximum depth. In the case of the \NMA, the cells with the biggest error, localized on the phenomena of the nominal flow condition are first to reach the limit. Progressively, the other cells with the refined fraction limits are the only one left to be refined. Due to this mechanism at the end of the adaptation, while the mesh size does not increase significantly, the average error is still reduced, resulting in a fast error convergence for the last adaptation iterations. This is of short duration since the mesh convergences fast toward its final state.

Since the \SMA\ adaptation process produces a mesh for every condition the error levels is always controlled and therefore the average error is lower than for the other methods. In fact it is impossible for the \MMA\ and \UR\ methods to do better since creating a unique mesh for different flow conditions will necessary lead to mesh size or error convergence inefficiencies as adapting for all conditions means that useless resolution is introduced for each condition and adapting on only one means that no resolution is dedicated to most conditions.

As previously discussed on the Burgers case, because both \SMA\ and \MMA\ methods manage to consistently reduce the error during the adaptation process the main influence for choosing one or the other comes from the cost to reach a specific error level. To this end, the ratio between the evaluation cost of the \MMA\ and \SMA\ methods (efficiency) has been represented against the target error level on figure \ref{fig:nmcomp_s}. It has been measured experimentally and modeled for different solver scaling using the method proposed in section \ref{sec:cost}. For both, the adaptation cost has been ignored ($\adcost=0$) since the remeshing, initialization and prolongation phases developed for this paper are not as efficiently implemented as the CFD solver \textit{elsA} making their comparison unrepresentative. Nevertheless, neglecting the adaptation cost is a conservative choice since it lowers \SMA's cost, without changing \MMA's. 

To reach a specific error level the \MMA\ mesh is always bigger than the \SMA. Indeed, as the shocks get more localized with the refinement a large area still needs to be refined in order to create a unique mean mesh, like in the central duct diverging section (see figure \ref{fig:local_err_mean_s}), whereas the \SMA\ method is able to keep refinement localized along the adaptation (see figure \ref{fig:local_err_det_s}). Therefore, the observed efficiency comes from the difference between solving the CFD problem once per evaluation on a big mesh (\MMA) and the cost of adaptation, several solves, for each condition on smaller meshes (\SMA). 

With the model, a strong dependency of the efficiency to the solver scaling parameter is observed. For an optimal solver scaling ($\cfdscaling=1$) \MMA\ is expected to always be more efficient than \SMA. Increasing it reduces the \MMA\ advantage zone to higher errors. This is mainly due to the bigger meshes used for \MMA\ evaluations. Indeed, with high  solver scaling, solving on big meshes gets comparatively more expensive than solving multiple times on small meshes.

In the experiments, \MMA\ is more efficient than \SMA\ for high errors targets until $4.10^{-3}$. This is consistent to what's modeled using solver scaling between $1.5$ and 2. Nevertheless, for high error targets the observed efficiency is almost double of the modeled one. This comes from the fact that the model only consider the cost of solving the CFD problem but do not consider any overhead costs like mesh partitioning costs that are inherent to the CFD solver. These overheads benefit to the \MMA\ efficiency because they are only computed once with \MMA\  while they are repeated for each \SMA\ iteration. As the error target gets lower the solving cost gets higher, therefore, these overheads become negligible and the costs converge to the same values.

To summarize, when compared to \SMA, \MMA\ should be used when targeting reasonable error levels. The specific error level from which it becomes more efficient to use \SMA\ is mainly depending on the flow topology variability, as it will set the mean mesh size for a given error level. The solver scaling is also an important parameter as poorly scaling solvers will reduce \MMA\ efficiency. Finally, adaptation and solver overhead should also be considered as their positive impact on \MMA\ efficiency is non negligible for moderate error levels.

\subsubsection{Detailed error analysis}

\begin{figure}[htb!]
	\centering
	\includegraphics[width=0.75\textwidth]{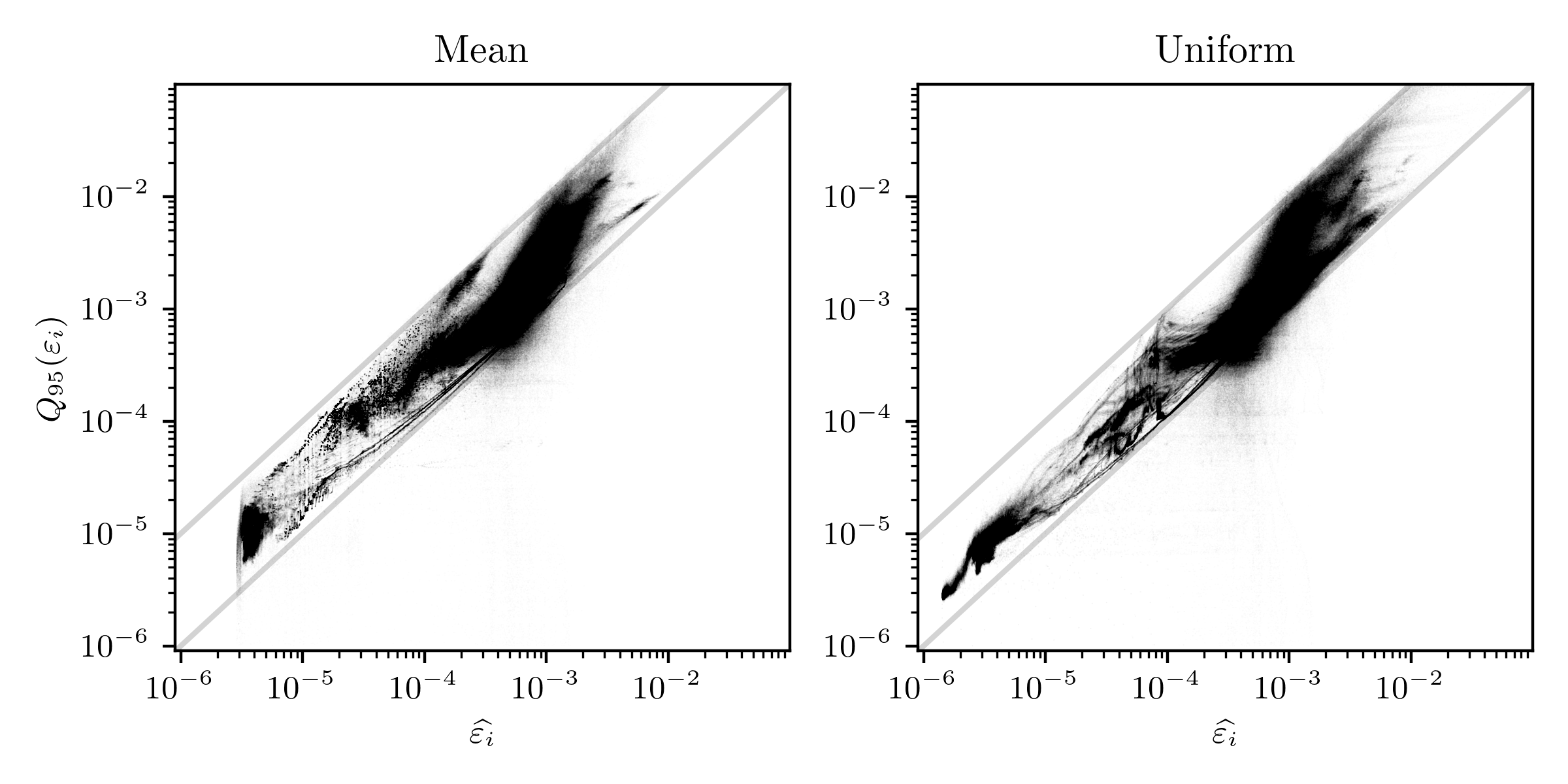}
	\caption{95\% quantile of the local error with respect to the corresponding average value. For a \MMA\ grid (left) and a uniform grid (right) achieving the same average error target (for \MMA\ $\reffrac=0.3$, $\constrsize=20$ and quadrature is MC).}\label{fig:error_clouds_comp}
\end{figure}

The \MMA\  process as it is implemented here is aimed at minimizing the estimated average error. Nevertheless, since we want to build a mesh adapted for a continuous range of parameters this could be insufficient in some cases. Indeed, while the error is guaranteed to be improved on average, some outliers could see higher local error levels. To check that \MMA\ does not exhibit such behaviors, a 95\% quantile of the measured error relative to the nominal inlet Mach number is computed. It is displayed with respect to the corresponding average error in each cells of the domain, for the \MMA\ (left) and a equivalent uniform grid (depth 5) (right) on figure \ref{fig:error_clouds_comp}.

For both the \MMA\ and the \UR\ mesh, the quantile remains confined within a band of one order of magnitude above the mean value. Therefore, the local refinement choices of the \MMA\ method do not induce higher error outliers than an uniform refinement. This can be explained by the fact that for each quadrature condition used to build the mean mesh, the refinement is in general wider than the targeted features mainly because of smoothing and the high refined fraction chosen. Consequently, in most of the domain the refined area is continuous (see \ref{fig:local_err_mean_s}) and almost every condition gets sufficient refinement where needed.

\begin{figure}[htb!]
    \begin{subfigure}[t]{0.67\linewidth}
	\centering
    \includegraphics[]{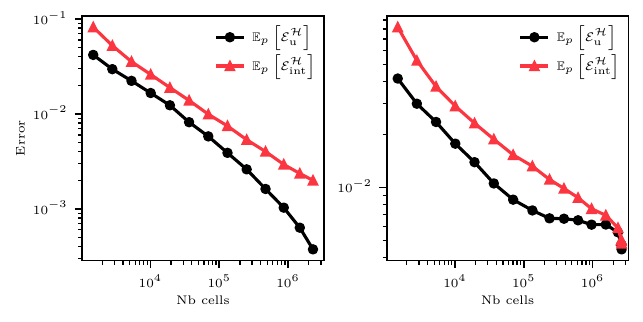}
	\caption{Evolution of the  true average error and it's estimation with the interpolation error for the \MMA\ (left) and \NMA\ (right) methods.}\label{fig:comp_err_itrp_mean_s}
\end{subfigure}
    \begin{subfigure}[t]{0.33\linewidth}
    \centering
     \includegraphics[]{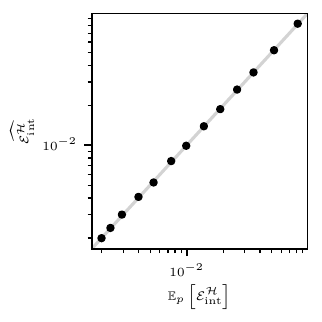}
	\caption{Comparison of the average interpolation error and its estimation on the quadrature conditions.}\label{fig:itrpcscvs}
    \end{subfigure}
    \caption{Evaluation of the average error estimation (with $\reffrac=0.3$, $\constrsize=20$ and MC quadrature).}
\end{figure}

The error measurement method used for the scramjet configuration has several limitations. First, it is expensive to compute, because the flow for each of the validation samples needs to be computed on the reference grid of 6.6M cells and also at every step of the adaptations process to evaluate. Then, the fine reference mesh allows for small but not negligible errors, meaning that convergence of the proposed error measurement does not mean convergence toward the true error. Finally, the reference limits the refinement depth of the adaptations and therefore their ability to reduce the error.
To prevent these issues, the use of interpolation error as an error convergence estimator is investigated. On the figure \ref{fig:comp_err_itrp_mean_s} the errors using the reference solution and the interpolation error estimate have been compared for the \MMA\ and \NMA\ adaptation methods.

The interpolation error estimation is an upper bound of the error measured from the reference. This is the preferred behavior for this kind of estimates as one would rather have the error target guaranteed at the expense of a more refined mean mesh. In addition, the difference increases at the end of the adaptation process as the interpolation error keeps a constant convergence order while the true error convergence accelerates. This divergence is due to the fact that the observed solution tends toward the reference solution which is yet not the analytical one, as the cells get refined to a depth of 6. Finally, the good estimate made by the interpolation estimator means that during the \MMA\ process the true error is on average dominated by interpolation errors. Therefore, the interpolation error can be used as an average error convergence estimator for the \MMA\ method instead of the expensive and reference sensitive error measurement used before.

For the \NMA\ adaptation, while both error estimators start with comparable behaviors, at the end of the adaptation process the measurement against reference highlights a plateau that is not captured by the interpolation error estimator. Therefore, the interpolation error is not the dominant error source at the end of the \NMA\ adaptation process. This means that the error mainly comes from other sources. Thus, in addition to not being well resolved, the features far from the nominal condition are also misplaced and/or misrepresented. In this case,  the mesh depth still increases in some areas of interest, effectively reducing the average interpolation error which therefore does not feature the plateau that the misrepresentation of most features should impose.

Finally, the same conclusion that has been made on the Burgers case is verified: in the the presence of conditional variability an adaptation method that fails to refine all the areas of interest, such as the \NMA\ adaptation , does not guarantee dominance of the interpolation error on average and thus the effectiveness of the interpolation error as an error convergence estimator.

The \MMA\ method has shown to be robust enough to make the interpolation error estimator a viable option for convergence evaluation. Now, to evaluate the possibility of assessing the convergence during the adaptation process without additional computation, the interpolation estimation of the error on the quadrature and the validation set have been compared on figure \ref{fig:itrpcscvs}. A good agreement between both approach is found during the whole adaptation process. This validates the behavior observed with the Burgers case and consolidates the confidence one can have in estimating the error with this technique.

%% file: main.bbl
\begin{thebibliography}{37}
\expandafter\ifx\csname natexlab\endcsname\relax\def\natexlab#1{#1}\fi
\providecommand{\url}[1]{\texttt{#1}}
\providecommand{\href}[2]{#2}
\providecommand{\path}[1]{#1}
\providecommand{\DOIprefix}{doi:}
\providecommand{\ArXivprefix}{arXiv:}
\providecommand{\URLprefix}{URL: }
\providecommand{\Pubmedprefix}{pmid:}
\providecommand{\doi}[1]{\href{http://dx.doi.org/#1}{\path{#1}}}
\providecommand{\Pubmed}[1]{\href{pmid:#1}{\path{#1}}}
\providecommand{\bibinfo}[2]{#2}
\ifx\xfnm\relax \def\xfnm[#1]{\unskip,\space#1}\fi
\bibitem[{Alauzet and Loseille(2016)}]{alauzet&loseille}
\bibinfo{author}{F.~Alauzet}, \bibinfo{author}{A.~Loseille},
\newblock \bibinfo{title}{A decade of progress on anisotropic mesh adaptation
  for computational fluid dynamics},
\newblock \bibinfo{journal}{Computer-Aided Design} \bibinfo{volume}{72}
  (\bibinfo{year}{2016}) \bibinfo{pages}{13--39}.
\bibitem[{Habashi et~al.(2000)Habashi, Dompierre, Bourgault, {Ait-Ali-Yahia},
  Fortin, and Vallet}]{Habashi2000}
\bibinfo{author}{W.~G. Habashi}, \bibinfo{author}{J.~Dompierre},
  \bibinfo{author}{Y.~Bourgault}, \bibinfo{author}{D.~{Ait-Ali-Yahia}},
  \bibinfo{author}{M.~Fortin}, \bibinfo{author}{M.-G. Vallet},
\newblock \bibinfo{title}{Anisotropic mesh adaptation: Towards
  user-independent, mesh-independent and solver-independent {{CFD}}. {{Part
  I}}: General principles},
\newblock \bibinfo{journal}{International Journal for Numerical Methods in
  Fluids} \bibinfo{volume}{32} (\bibinfo{year}{2000})
  \bibinfo{pages}{725--744}.
  \DOIprefix\doi{10.1002/(SICI)1097-0363(20000330)32:6<725::AID-FLD935>3.0.CO;2-4}.
\bibitem[{Baker(1997)}]{Baker1997}
\bibinfo{author}{T.~J. Baker},
\newblock \bibinfo{title}{Mesh adaptation strategies for problems in fluid
  dynamics},
\newblock \bibinfo{journal}{Finite Elements in Analysis and Design}
  \bibinfo{volume}{25} (\bibinfo{year}{1997}) \bibinfo{pages}{243--273}.
\bibitem[{Alauzet et~al.(2022)Alauzet, Frazza, and
  Papadogiannis}]{alauzet_turbo}
\bibinfo{author}{F.~Alauzet}, \bibinfo{author}{L.~Frazza},
  \bibinfo{author}{D.~Papadogiannis},
\newblock \bibinfo{title}{Periodic adjoints and anisotropic mesh adaptation in
  rotating frame for high-fidelity {{RANS}} turbomachinery applications},
\newblock \bibinfo{journal}{Journal of Computational Physics}
  \bibinfo{volume}{450} (\bibinfo{year}{2022}) \bibinfo{pages}{110814}.
\bibitem[{Vivarelli et~al.(2018)Vivarelli, Qin, Shahpar, and
  Radford}]{Vivarelli2018}
\bibinfo{author}{G.~Vivarelli}, \bibinfo{author}{N.~Qin},
  \bibinfo{author}{S.~Shahpar}, \bibinfo{author}{D.~Radford},
\newblock \bibinfo{title}{Efficient adjoint-based mesh adaptation applied to
  turbo-machinery flows},
\newblock in: \bibinfo{booktitle}{Proceedings of {{ASME Turbo Expo}} 2018},
  \bibinfo{year}{2018}.
\bibitem[{Van~Langenhove et~al.(2018)Van~Langenhove, Lucor, Alauzet, and
  Belme}]{Langenhove2018}
\bibinfo{author}{J.~Van~Langenhove}, \bibinfo{author}{D.~Lucor},
  \bibinfo{author}{F.~Alauzet}, \bibinfo{author}{A.~Belme},
\newblock \bibinfo{title}{Goal-oriented error control of stochastic system
  approximations using metric-based anisotropic adaptations},
\newblock \bibinfo{journal}{Journal of Computational Physics}
  \bibinfo{volume}{374} (\bibinfo{year}{2018}) \bibinfo{pages}{384--412}.
  \DOIprefix\doi{10.1016/j.jcp.2018.07.044}.
\bibitem[{Capriati et~al.(2022)Capriati, Cortesi, Magin, and
  Congedo}]{Capriati2022}
\bibinfo{author}{M.~Capriati}, \bibinfo{author}{A.~Cortesi},
  \bibinfo{author}{T.~E. Magin}, \bibinfo{author}{P.~M. Congedo},
\newblock \bibinfo{title}{Stagnation point heat flux characterization under
  numerical error and boundary conditions uncertainty},
\newblock \bibinfo{journal}{European Journal of Mechanics / B Fluids}
  \bibinfo{volume}{95} (\bibinfo{year}{2022}) \bibinfo{pages}{221--230}.
\bibitem[{Palacios et~al.(2012)Palacios, Duraisamy, Alonso, and
  Zuazua}]{Palacios2012}
\bibinfo{author}{F.~Palacios}, \bibinfo{author}{K.~Duraisamy},
  \bibinfo{author}{J.~J. Alonso}, \bibinfo{author}{E.~Zuazua},
\newblock \bibinfo{title}{Robust {{Grid Adaptation}} for {{Efficient
  Uncertainty Quantification}}},
\newblock \bibinfo{journal}{AIAA Journal} \bibinfo{volume}{50}
  (\bibinfo{year}{2012}) \bibinfo{pages}{1538--1546}.
  \DOIprefix\doi{10.2514/1.J051379}.
\bibitem[{Barral et~al.(2024)Barral, Taddei, and Tifouti}]{Barral}
\bibinfo{author}{N.~Barral}, \bibinfo{author}{T.~Taddei},
  \bibinfo{author}{I.~Tifouti},
\newblock \bibinfo{title}{Registration-based model reduction of parameterized
  {{PDEs}} with spatio-parameter adaptivity},
\newblock \bibinfo{journal}{Journal of Computational Physics}
  \bibinfo{volume}{499} (\bibinfo{year}{2024}) \bibinfo{pages}{112727}.
  \DOIprefix\doi{10.1016/j.jcp.2023.112727}.
\bibitem[{Peraire et~al.(1987)Peraire, Vahdati, Morgan, and
  Zienkiewicz}]{Peraire1987}
\bibinfo{author}{J.~Peraire}, \bibinfo{author}{M.~Vahdati},
  \bibinfo{author}{K.~Morgan}, \bibinfo{author}{O.~Zienkiewicz},
\newblock \bibinfo{title}{Adaptive remeshing for compressible flow
  computations},
\newblock \bibinfo{journal}{Journal of Computational Physics}
  \bibinfo{volume}{72} (\bibinfo{year}{1987}) \bibinfo{pages}{449--466}.
  \DOIprefix\doi{10.1016/0021-9991(87)90093-3}.
\bibitem[{Benard et~al.(2016)Benard, Balarac, Moureau, Dobrzynski, Lartigue,
  and {d'Angelo}}]{Benard&al}
\bibinfo{author}{P.~Benard}, \bibinfo{author}{G.~Balarac},
  \bibinfo{author}{V.~Moureau}, \bibinfo{author}{C.~Dobrzynski},
  \bibinfo{author}{G.~Lartigue}, \bibinfo{author}{Y.~{d'Angelo}},
\newblock \bibinfo{title}{Mesh adaptation for large-eddy simulations in complex
  geometries},
\newblock \bibinfo{journal}{International Journal for Numerical Methods in
  Fluids, Wiley}  (\bibinfo{year}{2016}) \bibinfo{pages}{719--740}.
\bibitem[{Kumar(1981)}]{Kumar81}
\bibinfo{author}{A.~Kumar}, \bibinfo{title}{Numerical Analysis of the
  Scramjet-Inlet Flow Field by Using Two-Dimensional {{Navier-Stokes}}
  Equations}, \bibinfo{type}{Technical {{Report}}}
  \bibinfo{number}{NASA-TP-1940}, NASA Langley Research Center,
  \bibinfo{year}{1981}.
\bibitem[{{Castro-D{\'i}az} et~al.(1997){Castro-D{\'i}az}, Hecht, Mohammadi,
  and Pironneau}]{CastroDiaz1997}
\bibinfo{author}{M.~J. {Castro-D{\'i}az}}, \bibinfo{author}{F.~Hecht},
  \bibinfo{author}{B.~Mohammadi}, \bibinfo{author}{O.~Pironneau},
\newblock \bibinfo{title}{Anisotropic unstructured mesh adaption for flow
  simulations},
\newblock \bibinfo{journal}{International Journal for Numerical Methods in
  Fluids} \bibinfo{volume}{25} (\bibinfo{year}{1997})
  \bibinfo{pages}{475--491}.
  \DOIprefix\doi{10.1002/(SICI)1097-0363(19970830)25:4<475::AID-FLD575>3.0.CO;2-6}.
\bibitem[{Alauset(2003)}]{Alauzet2003}
\bibinfo{author}{F.~Alauset}, \bibinfo{title}{Adaptation de Maillage Anisotrope
  En Trois Dimensions. {{Application}} Aux Simulations Instationnaires En
  {{M{\'e}canique}} Des {{Fluides}}.}, Ph.D. thesis, Montpellier II,
  \bibinfo{year}{2003}.
\bibitem[{Loseille et~al.(2007)Loseille, Dervieux, Frey, and
  Alauzet}]{Loseille&al}
\bibinfo{author}{A.~Loseille}, \bibinfo{author}{A.~Dervieux},
  \bibinfo{author}{P.~Frey}, \bibinfo{author}{F.~Alauzet},
\newblock \bibinfo{title}{Achievement of global second order mesh convergence
  for discontinuous flows with adapted unstructured meshes},
\newblock \bibinfo{journal}{American Institute of Aeronautics and Astronautics}
   (\bibinfo{year}{2007}).
\bibitem[{Gnoffo(1982)}]{Gnoffo1982}
\bibinfo{author}{P.~A. Gnoffo},
\newblock \bibinfo{title}{A vectorized, finite-volume, adaptive-grid algorithm
  for {{Navier-Stokes}} calculations},
\newblock \bibinfo{journal}{Applied Mathematics and Computation}
  \bibinfo{volume}{10--11} (\bibinfo{year}{1982}) \bibinfo{pages}{819--835}.
  \DOIprefix\doi{10.1016/0096-3003(82)90224-7}.
\bibitem[{Berger and Jameson(1985)}]{Berger1985}
\bibinfo{author}{M.~J. Berger}, \bibinfo{author}{A.~Jameson},
\newblock \bibinfo{title}{Automatic adaptive grid refinement for the {{Euler}}
  equations},
\newblock \bibinfo{journal}{AIAA Journal} \bibinfo{volume}{23}
  (\bibinfo{year}{1985}) \bibinfo{pages}{561--568}.
  \DOIprefix\doi{10.2514/3.8951}.
\bibitem[{Kompenhans et~al.(2016)Kompenhans, Rubio, Ferrer, and
  Valero}]{Kompenhans2016}
\bibinfo{author}{M.~Kompenhans}, \bibinfo{author}{G.~Rubio},
  \bibinfo{author}{E.~Ferrer}, \bibinfo{author}{E.~Valero},
\newblock \bibinfo{title}{Comparisons of p-adaptation strategies based on
  truncation- and discretisation-errors for high order discontinuous
  {{Galerkin}} methods},
\newblock \bibinfo{journal}{Computers \& Fluids} \bibinfo{volume}{139}
  (\bibinfo{year}{2016}) \bibinfo{pages}{36--46}.
  \DOIprefix\doi{10.1016/j.compfluid.2016.03.026}.
\bibitem[{Lovely and Haimes(1999)}]{Lovely&Haimes}
\bibinfo{author}{D.~Lovely}, \bibinfo{author}{R.~Haimes},
\newblock \bibinfo{title}{Shock detection from computational fluid dynamics
  results},
\newblock in: \bibinfo{booktitle}{Proceedings of the 14th {{AIAA Computational
  Fluid Dynamics Conference}}}, \bibinfo{year}{1999}.
\bibitem[{Haimes and Kenwright(1999)}]{Haimes}
\bibinfo{author}{R.~Haimes}, \bibinfo{author}{D.~Kenwright},
\newblock \bibinfo{title}{On the velocity gradient tensor and fluid feature
  extraction},
\newblock \bibinfo{journal}{American Institute of Aeronautics and Astronautics}
   (\bibinfo{year}{1999}).
\bibitem[{Deck and Renard(2020)}]{deck}
\bibinfo{author}{S.~Deck}, \bibinfo{author}{N.~Renard},
\newblock \bibinfo{title}{Towards an enhanced protection of attached boundary
  layers in hybrid {{RANS}}/{{LES}} methods.},
\newblock \bibinfo{journal}{Journal of Computational Physics, Elsevier}
  (\bibinfo{year}{2020}).
\bibitem[{Zhou et~al.(2020)Zhou, Zhao, and Shi}]{Zhou&al}
\bibinfo{author}{L.~Zhou}, \bibinfo{author}{R.~Zhao}, \bibinfo{author}{X.-P.
  Shi},
\newblock \bibinfo{title}{Entropy: Theory and new insights},
\newblock \bibinfo{publisher}{Ricardo Beltran-Chacon}, \bibinfo{year}{2020}.
\bibitem[{Liu et~al.(2019)Liu, Gao, Dong, Wang, Liu, Zhang, Cai, and
  Gui}]{Liu2019}
\bibinfo{author}{C.~Liu}, \bibinfo{author}{Y.-s. Gao}, \bibinfo{author}{X.-r.
  Dong}, \bibinfo{author}{Y.-q. Wang}, \bibinfo{author}{J.-m. Liu},
  \bibinfo{author}{Y.-n. Zhang}, \bibinfo{author}{X.-s. Cai},
  \bibinfo{author}{N.~Gui},
\newblock \bibinfo{title}{Third generation of vortex identification methods:
  {{Omega}} and {{Liutex}}/{{Rortex}} based systems},
\newblock \bibinfo{journal}{Journal of Hydrodynamics} \bibinfo{volume}{31}
  (\bibinfo{year}{2019}) \bibinfo{pages}{205--223}.
  \DOIprefix\doi{10.1007/s42241-019-0022-4}.
\bibitem[{Grenouilloux et~al.(2022)Grenouilloux, Balarac, Leparoux, Moureau,
  Lartigue, B{\'e}nard, Mercier, and Ferrey}]{Grenouilloux2022}
\bibinfo{author}{A.~Grenouilloux}, \bibinfo{author}{G.~Balarac},
  \bibinfo{author}{J.~Leparoux}, \bibinfo{author}{V.~Moureau},
  \bibinfo{author}{G.~Lartigue}, \bibinfo{author}{P.~B{\'e}nard},
  \bibinfo{author}{R.~Mercier}, \bibinfo{author}{P.~Ferrey},
\newblock \bibinfo{title}{On the use of kinetic-energy balance for the
  feature-based mesh adaptation applied to large-eddy simulation in complex
  geometries},
\newblock in: \bibinfo{booktitle}{Proceedings of {{ASME Turbo Expo}} 2022},
  \bibinfo{year}{2022}.
\bibitem[{L{\"o}hner et~al.(1985)L{\"o}hner, Morgan, and
  Zienkiewicz}]{Lohner1985}
\bibinfo{author}{R.~L{\"o}hner}, \bibinfo{author}{K.~Morgan},
  \bibinfo{author}{O.~Zienkiewicz},
\newblock \bibinfo{title}{An adaptive finite element procedure for compressible
  high speed flows},
\newblock \bibinfo{journal}{Computer Methods in Applied Mechanics and
  Engineering} \bibinfo{volume}{51} (\bibinfo{year}{1985})
  \bibinfo{pages}{441--465}. \DOIprefix\doi{10.1016/0045-7825(85)90042-8}.
\bibitem[{Oden et~al.(1987)Oden, Strouboulis, and Devloo}]{Oden1987}
\bibinfo{author}{J.~T. Oden}, \bibinfo{author}{T.~Strouboulis},
  \bibinfo{author}{{\relax Ph}.~Devloo},
\newblock \bibinfo{title}{Adaptive finite element methods for high-speed
  compressible flows},
\newblock \bibinfo{journal}{International Journal for Numerical Methods in
  Fluids} \bibinfo{volume}{7} (\bibinfo{year}{1987})
  \bibinfo{pages}{1211--1228}. \DOIprefix\doi{10.1002/fld.1650071105}.
\bibitem[{Wu et~al.(1990)Wu, Zhu, Szmelter, and Zienkiewicz}]{Wu1990}
\bibinfo{author}{J.~Wu}, \bibinfo{author}{J.~Z. Zhu},
  \bibinfo{author}{J.~Szmelter}, \bibinfo{author}{O.~C. Zienkiewicz},
\newblock \bibinfo{title}{Error estimation and adaptivity in {{Navier-Stokes}}
  incompressible flows},
\newblock \bibinfo{journal}{Computational Mechanics} \bibinfo{volume}{6}
  (\bibinfo{year}{1990}) \bibinfo{pages}{259--270}.
  \DOIprefix\doi{10.1007/BF00370106}.
\bibitem[{Giles and Pierce(1999)}]{Giles1999}
\bibinfo{author}{M.~Giles}, \bibinfo{author}{N.~Pierce},
\newblock \bibinfo{title}{Improved lift and drag estimates using adjoint
  {{Euler}} equations},
\newblock in: \bibinfo{booktitle}{14th {{Computational Fluid Dynamics
  Conference}}}, \bibinfo{publisher}{{American Institute of Aeronautics and
  Astronautics}}, \bibinfo{address}{Norfolk,VA,U.S.A.}, \bibinfo{year}{1999}.
  \DOIprefix\doi{10.2514/6.1999-3293}.
\bibitem[{Venditti and Darmofal(2003)}]{Venditti2003}
\bibinfo{author}{D.~A. Venditti}, \bibinfo{author}{D.~L. Darmofal},
\newblock \bibinfo{title}{Anisotropic grid adaptation for functional outputs:
  Application to two-dimensional viscous flows},
\newblock \bibinfo{journal}{Journal of Computational Physics}
  \bibinfo{volume}{187} (\bibinfo{year}{2003}) \bibinfo{pages}{22--46}.
  \DOIprefix\doi{10.1016/S0021-9991(03)00074-3}.
\bibitem[{Benoit et~al.(2015)Benoit, P{\'e}ron, and Landier}]{cassio2015}
\bibinfo{author}{C.~Benoit}, \bibinfo{author}{S.~P{\'e}ron},
  \bibinfo{author}{S.~Landier},
\newblock \bibinfo{title}{Cassiopee: A {{CFD}} pre- and post-processing tool},
\newblock \bibinfo{journal}{Aerospace Science and Technology}
  \bibinfo{volume}{45} (\bibinfo{year}{2015}) \bibinfo{pages}{272--283}.
\bibitem[{Abramowitz and Stegun(2013)}]{Abramowitz}
\bibinfo{editor}{M.~Abramowitz}, \bibinfo{editor}{I.~A. Stegun} (Eds.),
  \bibinfo{title}{Handbook of Mathematical Functions: With Formulas, Graphs,
  and Mathematical Tables}, Dover Books on Mathematics, \bibinfo{edition}{9.
  dover print.; [nachdr. der ausg. von 1972]} ed., \bibinfo{publisher}{Dover
  Publ}, \bibinfo{address}{New York, NY}, \bibinfo{year}{2013}.
\bibitem[{Liu(2004)}]{Liu2004}
\bibinfo{author}{J.~S. Liu}, \bibinfo{title}{Monte {{Carlo Strategies}} in
  {{Scientific Computing}}}, Springer {{Series}} in {{Statistics}},
  \bibinfo{publisher}{Springer New York}, \bibinfo{address}{New York, NY},
  \bibinfo{year}{2004}. \DOIprefix\doi{10.1007/978-0-387-76371-2}.
\bibitem[{Hopf(1950)}]{Hopf1950}
\bibinfo{author}{E.~Hopf},
\newblock \bibinfo{title}{The partial differential equation ut + uux =
  {$M$}xx},
\newblock \bibinfo{journal}{Communications on Pure and Applied Mathematics}
  \bibinfo{volume}{3} (\bibinfo{year}{1950}) \bibinfo{pages}{201--230}.
  \DOIprefix\doi{10.1002/cpa.3160030302}.
\bibitem[{Roe(1981)}]{Roe1981}
\bibinfo{author}{P.~Roe},
\newblock \bibinfo{title}{Approximate {{Riemann}} solvers, parameter vectors,
  and difference schemes},
\newblock \bibinfo{journal}{Journal of Computational Physics}
  \bibinfo{volume}{43} (\bibinfo{year}{1981}) \bibinfo{pages}{357--372}.
  \DOIprefix\doi{10.1016/0021-9991(81)90128-5}.
\bibitem[{Balay et~al.(2023)Balay, Abhyankar, Adams, Benson, and Brown}]{petsc}
\bibinfo{author}{S.~Balay}, \bibinfo{author}{S.~Abhyankar},
  \bibinfo{author}{M.~Adams}, \bibinfo{author}{S.~Benson},
  \bibinfo{author}{J.~Brown}, \bibinfo{title}{{{PETSc}}/{{TAO Users Manual}}},
  \bibinfo{type}{Technical Report} \bibinfo{number}{ANL-21/39-Rev.3.19,
  1968587, 181821}, \bibinfo{year}{2023}. \DOIprefix\doi{10.2172/1968587}.
\bibitem[{Van~Langenhove(2017)}]{Langenhove2017}
\bibinfo{author}{J.~Van~Langenhove}, \bibinfo{title}{Adaptive Control of
  Deterministic and Stochastic Approximation Errors in Simulations of
  Compressible Flow}, Ph.D. thesis, Universit{\'e} Pierre et Marie Curie -
  Paris VI, \bibinfo{address}{Paris}, \bibinfo{year}{2017}.
\bibitem[{Cambier et~al.(2013)Cambier, Heib, and Plot}]{elsa2016}
\bibinfo{author}{L.~Cambier}, \bibinfo{author}{S.~Heib},
  \bibinfo{author}{S.~Plot},
\newblock \bibinfo{title}{The {{ONERA elsA CFD}} software : Input from research
  and feedback from industry},
\newblock \bibinfo{journal}{Mechanics \& Industry, EDP Sciences}
  \bibinfo{volume}{14} (\bibinfo{year}{2013}) \bibinfo{pages}{159--174}.

\end{thebibliography}
